\documentstyle[aps,preprint,epsf]{revtex}
\begin{document}
\preprint{UW/PT-97-19}
\draft 
\title{Effective non-perturbative real-time dynamics of soft modes in 
  hot gauge theories}
\author{D.T.~Son} 
\address{Department of Physics, University of
  Washington, Seattle WA 98195} 
\date{July 1997} 
\maketitle
\begin{abstract}
  We derive, from first principle, the Fokker-Planck equation
  describing the non-perturbative real-time evolution of a gauge field
  at high temperatures on the spatial scale $(g^2T)^{-1}$ and the time
  scale $(g^4T)^{-1}$, where $g$ is the gauge coupling and $T$ is the
  temperature.  The knowledge of the effective dynamics on such
  spatial and time scales is crucial for the understanding and
  quantitative calculation of the baryon number violation rate at high
  temperatures in the electroweak theory.  The Fokker-Planck equation,
  which describes the diffusive motion of soft components of the gauge
  field in the space of field configurations, is shown to be
  gauge-invariant and consistent with the known thermodynamics of soft
  gauge modes.  We discuss the ways the Fokker-Planck equation can be
  made useful for numerical simulations of long-distance dynamics of
  hot gauge theories.
\end{abstract}

\newpage

\section{Introduction}

Despite a relatively long history, the problem of computing the rate
of baryon number violation in the symmetric phase of the electroweak
model still does not have the ultimate solution.  Conventionally
assumed to be (parametrically) of order $O(\alpha_W^4T^4)$, where $T$
is the temperature, this rate has been re-evaluated to be suppressed
further by an additional factor of $\alpha_W$ due to damping phenomena
in hot plasma, which bring the rate down to $O(\alpha_W^5T^4)$
\cite{ASY,HuetSon}.  While the numerical coefficient of $\alpha_W^5
T^4$ is still not found, its knowledge is crucial for computing the
baryon number produced by electroweak baryogenesis.  

The physics underlying $B$ violation is sensitive to a particular type
of modes in hot plasma: the almost static, soft magnetic modes with
spatial momentum of order $g^2T$.\footnote{Hereafter, by ``soft''
  modes we will understand these modes.  We assume the gauge coupling
  at the energy scale $T$ to be arbitrarily small, $g(T)\ll1$.} The
same modes are well known to be the ones that cause the breakdown of
perturbation theory at high temperatures.  This flaw of the
perturbation theory for hot gauge plasma has been viewed mostly from
the context of static (equilibrium) thermal field theory, where the
perturbative expansion of any equilibrium quantity, like the free
energy, breaks down at the order that receives substantial
contribution from static, soft $g^2T$ magnetic modes.  The process of
$B$ violation at high temperatures presents another side of the
complications due to soft magnetic modes: their real-time {\em
  dynamics} is also non-perturbative.

For static characteristics, techniques have been develop to isolate
and, in principle, compute the contributions from $g^2T$ modes to
physical quantities.  In one of the approaches, the
dimensional-reduction technique\cite{Shaposhnikov}, all hard,
perturbative modes are integrated out analytically, and the theory for
the remaining soft modes is essentially a 3d classical gauge theory at
finite temperature.  The problem of computing non-perturbative
contributions to any physical quantity is reduced to solving the
latter.

In contrast, our knowledge of how to deal with the real-time dynamics
of $g^2T$ modes in hot gauge plasma is much less reliable.  It has
been suggested that simulating soft modes in a classical theory can
solve the problem in the quantum
theory\cite{AmbjornKrasnitz,TangSmit}.  The justification for using a
classical theory is the large occupation numbers of soft modes: when
many bosons are on the same energy level, there is little difference
between quantum and classical theories.  Unfortunately, this method
does not work because in a gauge theory, the dynamics of the soft
modes depend crucially on the hard ones \cite{ASY}.  Even for
classical theories, this dependence is present: the soft dynamics is
different in classical theories with different lattice
cutoffs\cite{MooreTurok}.  The situation is in sharp contrast with the
static case, where the hard sector only slightly (by an amount small
in the weak-coupling limit) modifies the classical Hamiltonian of the
soft one.

At present time, it seems that the only way one can try to treat the
real-time dynamics of soft modes is to ``integrate out'' the hard
modes and see what is left out for the soft ones.  The procedure
should resemble Wilsonian approach to renormalization group.

Integrating out hard modes in the real-time framework is not something
new: the general formalism suitable for performing such integration,
the ``influence functional'' approach, has been developed for quite a
while \cite{Feynman,CalzettaHu}.  Recently, the same method has been
applied to the scalar $\phi^4$ theory \cite{GreinerMuller}.  In the
latter case, the integration over hard modes leads to a stochastic
equation for the soft sector, in which a damping term and a stochastic
noise together keep the soft sector in thermal equilibrium.
Technically, however, the effective equation is rather complicated:
the noise and the damping are non-local in space and time, and both
depend in a non-trivial way on the soft field.  That the effective
soft dynamics is complicated in such a simple model as the $\phi^4$
theory is one of the reasons that the same calculation has not been
performed for the gauge field so far.

One the other hand, to gain some insight into the effective dynamics
of soft modes in hot gauge theories, the kinetic approach has been
explored \cite{Bodekeretal,HuetSon}.  In this approach, the soft
sector is described a classical field, while the hard sector is
replaced by a collection of classical particles moving on the soft
background.  Concerning the noise, it was suggested \cite{HuetSon}
that it can be naturally introduced to the kinetic approach by
considering the fluctuations of the distribution function of hard
particles.  The kinetic method yields a stochastic equation with
damping and noise terms \cite{HuetSon}.  It is remarkable that the
damping and the noise are local in time (they are still non-local in
space) and can be written is a relatively simple analytical form.  The
same is not true for the equation obtained in $\phi^4$ theory by the
influence functional method.

However, the use of the kinetic approach for hot gauge plasma is not
well justified for our problem.  While the kinetic equations are known
to correctly reproduce the physical quantities at $gT$ momentum scale
as the Debye screening length or the plasmon frequency
\cite{BlaizotIancu,Kelly}, it is not clear whether these equations
remain valid at the smaller scale of $g^2T$.  Moreover, in the
framework of this approach, the stochastic noise remains a
phenomenological quantity introduced by hand, and it is not obvious
that the noise is Gaussian as assumed in Ref.\ \cite{HuetSon} or it
has a more complicated distribution.  Moreover, it has not been
noticed in Ref.\ \cite{HuetSon} that there is an ambiguity in
understanding the Langevin equation.  Therefore, to reliably obtain
the effective dynamics of the soft modes, one has to re-derive the
equations from first principle, using a systematic approach.  The
influence functional technique seems to be the only available
candidate for the latter.

In this paper we present a systematic derivation of the soft dynamics
in hot gauge theories, that is based the influence functional method.
In this method, we divide the whole system in to a soft and a hard
sector, and integrate over all modes of the hard one.  Systematically
keeping track of the order of the coupling $g$, after long calculation
we recover the same set of Langevin equations as obtained from the
kinetic approach.

We furthermore observe that these equations are ambiguous, the fact
that has not been noticed in the original derivation of
\cite{HuetSon}.  By fixing this ambiguity, we derive the Fokker-Planck
equation that describes the diffusion in the space of field
configurations in a unique way.  We show that the Fokker-Planck
equation is gauge invariant, a crucial test for our ideology if one
takes into account that the operation of dividing the system into hard
and soft sectors does not respect gauge invariance.  We also point out
that the Fokker-Planck equation is consistent with the known
thermodynamics of soft modes, thus making a connection between the
dynamics of the soft modes and their statics (or thermodynamics).  The
Fokker-Planck equation (Eq.\ (\ref{FP_cont})) is the main result of
our paper.

This paper is organized as follows.  In Sec.\ \ref{sec:review} we
briefly review the general-purpose techniques that will be used in the
rest of the paper, namely, the Schwinger-Keldysh formalism, which
provides the general framework for real-time non-equilibrium field
theory, and the influence functional method which allows one to derive
the effective dynamics of a subsystem in a larger system.  In Sec.\ 
\ref{sec:core} we apply these techniques to the case of a hot pure
gauge theory, and find the effective dynamics of the soft modes in
this theory in the form of a Langevin equation.  We discuss the
ambiguity of the latter, and argue the way to fix this ambiguity.  We
then find the Fokker-Planck equation in the main text) that describes
the diffusion of the gauge field in the space of field configurations,
and check that, in the static limit, it leads to the same equilibrium
Gibbs distribution of the soft modes as expected from the static
method of dimensional reduction.  We also check the gauge invariance
of the Fokker-Planck equation.  Finally, we present the concluding
remarks in Sec.\ \ref{sec:conclusion}, where the relevancy of the
Fokker-Planck equation for numerical simulations of the plasma is also
discussed.  The Appendices contain various technical details.

\section{Brief review of Schwinger-Keldysh formalism and influence
functional technique}
\label{sec:review}

\subsection{Schwinger-Keldysh (Close-Time-Path) formalism}

The most general framework for dealing with field theories in a
real-time, non-equilibrium setting is the Schwinger-Keldysh, or
close-time-path (CTP) formalism.  In this section we will briefly
describe this formalism, mostly to introduce the notations that will
be used in the rest of the paper.  Detail treatments of the
Schwinger-Keldysh formalism can be found in Ref.\ 
\cite{LifshitzPitaevskii}.

Consider a quantum field theory of a generic field $\phi$ with some
action $S[\phi]$.  In the Schwinger-Keldysh approach, the theory is,
formally, put on a time contour which runs from some initial time
moment, say, $t=0$, to $t=+\infty$ and then goes back to $t=0$ (Fig.\ 
(\ref{figure1})).  The reason for having such a contour instead of the
simple one running from $t=-\infty$ to $t=+\infty$ is the need of
computing, in the real-time picture, the in-in matrix elements, i.e.\
those of a quantum operator sandwiched between two in-states instead
of, as in the standard treatment of scattering processes, the in-out
matrix elements between an in- and an out-state.  The field in the
upper and the lower parts of the contours need not to be the same, so
we will denote $\phi$ in the upper and lower parts of the contour as
$\phi_1$ and $\phi_2$, respectively, while reserving the notation
$\phi$ for the field on the whole contour.  The basic object in the
Schwinger-Keldysh formalism is the generating functional,
\begin{eqnarray}
  Z[J_1,J_2] & = &\int\!{\cal D}\phi_1\,{\cal D}\phi_2\,
  \rho[\phi_1(t=0),\phi_2(t=0)]\,\exp\biggl(
  iS[\phi_1]-iS[\phi_2]+
  i\int\!dx\,\Bigl(J_1(x)\phi_1(x)- \nonumber \\
  & &  -J_2(x)\phi_2(x)\Bigr)\biggr)
  \equiv\int\limits_C\!{\cal D}\phi\,\rho[\phi(t=0)]\,
  \exp(iS[\phi]+iJ\cdot\phi)
  \label{gen-func}
\end{eqnarray}
where $\rho[\phi_1,\phi_2]$ is the density matrix of the system at
$t=0$, $\rho[\phi_1,\phi_2]=\langle\phi_2|\hat{\rho}|\phi_1\rangle$
(we will take $\hat{\rho}$ to be that of thermal equilibrium),
$S[\phi]=S[\phi_1]-S[\phi_2]$, $J\cdot\phi=\int_C\!dx\,J(x)\phi(x)$,
and the index $C$ of the last integral in Eq.\ (\ref{gen-func}) means
that the variables in the integrations are taken on the contour $C$.
Further we will largely ignore the boundary term $\rho$ for simplicity
of notations when it is not essential.  All real-time physical
quantities can be extracted from the generating functional $Z$; for
example, the two-point correlation functions can be obtained by
differentiating $Z$ two times over $J$ at $J=0$,
\begin{equation}
  G(x,y)=-{\delta^2\over\delta J(x)\delta J(y)}\ln Z\biggl|_{J=0}=
  Z_0^{-1}\int\limits_C\!{\cal D}\phi\,\phi(x)\phi(y)
  \exp(iS[\phi])
  \label{Gxy_int}
\end{equation}
where $Z_0=\int_C\!{\cal D}\phi\,\text{e}^{iS[\phi]}$.  By putting $x$
and $y$ in different ways on the upper or the lower parts of the
contour, the generic notation $G$ corresponds to 4 different Green
functions, $G_{11}$, $G_{12}$, $G_{21}$, $G_{22}$, which in the
operator language can be written as follows
\begin{eqnarray}
  G_{11}(x,y) & = & \langle T\phi(x)\phi(y)\rangle,\qquad
  G_{12}(x,y)=\langle\phi(y)\phi(x)\rangle, \nonumber \\
  G_{22}(x,y) & = & \langle\bar{T}\phi(x)\phi(y)\rangle,\qquad
  G_{21}(x,y)=\langle\phi(x)\phi(y)\rangle,
  \label{G1122}
\end{eqnarray}
where now $\phi$ is the field operator, $T$ and $\bar{T}$ denote time
ordering and anti-ordering, respectively, and $\langle\cdots\rangle$
denotes the average over the thermal ensemble.  Therefore, Feynman,
anti-Feynman, and Wightman propagators can be recovered from $Z$.
This list exhausts all interesting two-point real-time correlation
functions.

The Schwinger-Keldysh formalism also includes a set of Feynman rules,
from which one can find any Green function in a weakly-coupled theory
by computing the relevant Feynman diagrams.  We will not go into
details here, see \cite{LifshitzPitaevskii}.

For illustration and further reference, let us write down the explicit
form of the thermal Green functions at equilibrium in some sample
theories.  The first example we will consider is the free massless
scalar field.  To compute the Green functions, it is convenient to
make use of their operator form, Eq.\ (\ref{G1122}). Expanding the
field operator in Fourier components,
\[
  \phi(t,x)=\int\!{d{\bf p}\over(2\pi)^32|{\bf p}|}\,
  (a_{\bf p}\text{e}^{-i|{\bf p}|t+i{\bf px}}+
   a^\dagger_{\bf p}\text{e}^{i|{\bf p}|t-i{\bf px}})
\]
where $p=|{\bf p}|$, and noticing that, in thermal equilibrium,
$\langle a^\dagger_{\bf p}a_{\bf p'}\rangle= (2\pi)^32|{\bf
  p}|\delta({\bf p}-{\bf p}')n_{\bf p}$, where $n_{\bf p}$ is the
Bose-Einstein distribution function, $n_{\bf p}=(\text{e}^{|{\bf
    p}|/T}-1)^{-1}$, we find the thermal Green functions for the free
massless scalar case in momentum representation,
\[
  G_{11}^s(\omega,{\bf p}) =  i\left(
  {1+n_{\bf p}\over\omega^2-{\bf p}^2+i\epsilon}-
  {n_{\bf p}\over\omega^2-{\bf p}^2-i\epsilon}\right)
\]
\[
  G_{22}^s(\omega,{\bf p}) =  i\left(
  {n_{\bf p}\over\omega^2-{\bf p}^2+i\epsilon}-
  {1+n_{\bf p}\over\omega^2-{\bf p}^2-i\epsilon}\right)
\]
\[
  G_{12}^s(\omega,{\bf p}) =  {\pi\over|{\bf p}|}
  \Bigl(n_{\bf p}\delta(\omega-|{\bf p}|)
  +(1+n_{\bf p})\delta(\omega+|{\bf p}|)\Bigr)
\]
\begin{equation}
  G_{21}^s(\omega,{\bf p}) =  {\pi\over|{\bf p}|}
  \Bigl((1+n_{\bf p})\delta(\omega-|{\bf p}|)
  +n_{\bf p}\delta(\omega+|{\bf p}|)\Bigr)
  \label{Gs}
\end{equation}
(the superscript $s$ means ``scalar'').  Let us notice a feature of
the Green functions that will turn out to be useful in further
discussion.  When $|{\bf p}|$ is much smaller than $T$, one has
$n_{\bf p}\gg 1$, and all four Green functions are approximately
equal.  $G^{11}\approx G^{12}\approx G^{21}\approx G^{22}$.  This fact
can be rephrased into the following statement: for soft modes (those
with $|{\bf p}|\ll T$), the path integral (\ref{Gxy_int}) is dominated
by such field configurations where the difference between the soft
components of the field in the upper and lower parts of the contour is
negligible.

As the second example, we will write down the photon propagator for
QED.  The most convenient gauge to work with, in thermal field theory,
is the Coulomb gauge where $\partial_iA^i=0$.  The photon propagator
contains, beside the two contour indices $\alpha$ and $\beta$ that can
be either 1 or 2, also two Lorentz indices.  The spatial components of
the photon propagator is proportional to the scalar propagator and has
the same (spatial) Lorentz structure as that of the transverse
projection operator,
\begin{equation}
  G_{\alpha\beta}^{ij}(\omega,{\bf p})=
  \left(\delta^{ij}-{p^ip^j\over{\bf p}^2}\right)
  G_{\alpha\beta}^s(\omega,{\bf p}),
  \qquad \alpha,\beta=1,2
  \label{G_photon_ij}
\end{equation}
The 00 component of the photon propagator is the same for all values
of the contour indices and does not have any singularity with respect
to $\omega$,
\begin{equation}
  G_{\alpha\beta}^{00}(\omega,{\bf p})={i\over{\bf p}^2}
  \qquad \alpha,\beta=1,2\
  \label{G_photon_00}
\end{equation}
which reflects the non-propagating nature of the field component
$A_0$.  All other components of $G$ vanish in the Coulomb gauge.

In a non-Abelian gauge theory with weak coupling, the propagator of
the gauge field, to the leading order of the coupling constant, is
given essentially by the same equations as in the QED case, Eqs.\ 
(\ref{G_photon_ij},\ref{G_photon_00}).  The propagator now has two
additional color indices $a$ and $b$, but the dependence over these
indices is trivial, i.e.\ via the Kronecker symbol $\delta^{ab}$.

\subsection{The influence functional formalism and the stochastic
  equation}

In order to find the dynamics of soft modes in gauge plasma, we will
divide the whole system into two subsystems, from which the first one
contains soft modes with spatial momentum of order $g^2T$, while the
second consists of hard modes.  We then integrate over all the degrees
of freedom in the second subsystem to obtain the effective theory for
the soft modes.  Before discussing the specific case of the gauge
theory, we need to review the general framework that allows one to
find the effective theory describing a subsystem of a larger system in
the context of real-time field theory.  Integration over hard modes is
usually performed in the Euclidean spacetime as, for example, in
dimensional reduction approach \cite{Shaposhnikov}.  Doing the same in
real-time, as we will see, is a little more complicated.  The
technique we will describe here is essentially the one developed and
used in Ref.\ \cite{CalzettaHu,GreinerMuller}, where further details
can be found.

Let us consider some field theory which contains two set of degrees of
freedom that will be generically denoted as $\phi$ and $\chi$.
Eventually, $\phi$ and $\chi$ will be identified with the soft and the
hard modes in the hot gauge plasma, but for the moment we will
regards them as just two subsystems making a full quantum system.
The action of the full system is a sum of those for the two subsystems
and an interaction term,
\[
  S[\phi,\chi]=S[\phi]+S[\chi]+S_{\text{int}}[\phi,\chi]
\]
We will try to find the effective theory of the $\phi$ field.  More
specifically, we will be interested in the $n$-point Green functions of
the $\phi$ field
\[
  G(x_1\ldots x_n)=Z_0^{-1}\int\limits_C\!{\cal D}\phi\,{\cal D}\chi\,
  \phi(x_1)\ldots\phi(x_n)\exp\Bigl(iS[\phi]+iS[\chi]+
  iS_{\text{int}}[\phi,\chi]\Bigr)
\]
where
\[
  Z_0=\int\limits_C\!{\cal D}\phi\,{\cal D}\chi\,
  \exp\Bigl(iS[\phi]+iS[\chi]+iS_{\text{int}}[\phi,\chi]\Bigr)
\]
and will derive a theory for $\phi$ alone, in which these Green
functions are the same as in the original full theory.

Take the integration over $\chi$, one finds
\begin{equation}
  G(x_1\ldots x_n)=Z_0^{-1}
  \int\limits_C\!{\cal D}\phi\,\phi(x_1)\ldots\phi(x_n)
  \exp\Bigl(iS[\phi]+iS_{\text{IF}}[\phi]\Bigr)
  \label{intdphi}
\end{equation}
where we have introduced the so-called influence functional
$S_{\text{{IF}}}$,
\begin{equation}
  \text{e}^{iS_{\text{IF}}[\phi]}=
  \int\limits_C\!{\cal D}\chi\,\exp\Bigl(iS[\chi]+
  iS_{\text{int}}[\phi,\chi]\Bigr)
  \label{SIF}
\end{equation}
or, in component notation,
\[
  \text{e}^{iS_{\text{IF}}[\phi_1,\phi_2]}=
  \int\!{\cal D}\chi_1\,{\cal D}\chi_2\,
  \exp\Bigl(iS[\chi_1]-iS[\chi_2]+iS_{\text{int}}
  [\phi_1,\phi_2;\chi_1,\chi_2]\Bigr)
\]
So far all our equations are exact.  Now we will make our first
approximation by assuming that the integral in Eq.\ (\ref{intdphi}) is
saturated by such configuration of $\phi$ where its values in the
upper and lower parts of the contour are close to each other,
\begin{equation}
  \phi_1=\phi+{\Delta\phi\over2},\qquad\phi_2=\phi-{\Delta\phi\over2},
  \label{phi12}
\end{equation}
where
\begin{equation}
  \Delta\phi\ll\phi
  \label{dplessp}
\end{equation}
In general, the condition (\ref{dplessp}) is not always satisfied.  We
make this assumption, keeping in mind the particular case of gauge
theories, where $\phi$ is the soft modes and $\chi$ represent hard
ones.  When the formalism is applied to the hot gauge theories, we
will verify explicitly that (\ref{dplessp}) is correct.  However, not
to claim any rigorousness, this assumption can be understood in the
following simple way.  Let us recall the field $\phi$ will be
identified with the soft modes in plasma.  We have seen (see Sec.\ 
\ref{sec:review}) that for the free scalar field the path integral is
saturated by field configurations having the soft components on the
two parts of the contour close to each other.  Eq.\ (\ref{dplessp})
means that the same is valid for the soft components of the gauge
field in the interacting theory.  Again, this is not a rigorous proof
and one needs to verify Eq.\ (\ref{dplessp}) specifically for the case
of hot gauge theory.

Though $\phi_1\approx\phi_2$, one cannot simply put $\phi_1=\phi_2$ to
the exponent in Eq.\ (\ref{intdphi}), since the exponential vanishes
if $\phi_1=\phi_2$.  In fact, one can see that when $\Delta\phi=0$, or
$\phi_1=\phi_2$, $S[\phi]$ vanishes: the contributions to $S[\phi]$
from the upper part and the lower part of the contour cancel each
other.  From unitarity it can be shown that $S_{\text{IF}}[\phi]$ also
vanishes when $\Delta\phi=0$.  So, one should expand the exponent over
$\Delta\phi$, keeping the latter small but finite.  For reasons that
will be clear in subsequent discussions, we will be interested in
terms of order $\Delta\phi$ and $(\Delta\phi)^2$, but not
$(\Delta\phi)^3$ and higher.  With the symmetric definition of $\phi$
as in Eq.\ (\ref{phi12}), the expansion of $S[\phi]$, up to the
$(\Delta\phi)^2$ term inclusively, is
\begin{equation}
  S[\phi]=S[\phi_1]-S[\phi_2]=\int\!dx\,
  {\partial S\over\partial\phi(x)}\Delta\phi(x)+
  O\Bigl((\Delta\phi)^3\Bigr)
  \label{S=S1-S2}
\end{equation}
The most general expression one may obtain expanding
$S_{\text{IF}}[\phi]$ over $\Delta\phi$, up to the order
$(\Delta\phi)^2$, is
\[
  S_{\text{IF}}[\phi]=-\int\!dx\,j(x)\Delta\phi(x)+
  {i\over2}\int\!dx\,dy\,\Delta\phi(x)N(x,y)\Delta\phi(y)
\]
(the time integration here, as well as in Eq.\ (\ref{S=S1-S2}), is
along the Minkowskian time axis but not the time contour $C$) where
we have introduced the following notations
\begin{equation}
  j(x)=-{1\over2}\left({\delta S_{\text{IF}}\over\delta\phi_1(x)}+
  {\delta S_{\text{IF}}\over\delta\phi_2(x)}\right)
  \label{j}
\end{equation}
and
\begin{equation}
  N(x,y)=-{i\over4}\left(
  {\delta^2S_{\text{IF}}\over\delta\phi_1(x)\delta\phi_1(y)}+
  {\delta^2S_{\text{IF}}\over\delta\phi_1(x)\delta\phi_2(y)}+
  {\delta^2S_{\text{IF}}\over\delta\phi_2(x)\delta\phi_1(y)}+
  {\delta^2S_{\text{IF}}\over\delta\phi_2(x)\delta\phi_2(y)}\right)
  \label{N}
\end{equation}
Our conventions are chosen in such a way that in the case of gauge
fields $j$ and $N$ are both real.  Note that $j$ and $N$ are,
generally speaking, functionals depending on $\phi$.  

Now the correlation function can be written in the following form,
\begin{eqnarray}
  G(x_1\ldots x_n) & = & Z_0^{-1}\int\!{\cal D}\phi\,{\cal D}\Delta\phi\,
  \phi(x_1)\ldots\phi(x_n)\times \nonumber \\
  & & \times\exp\left(i\int\!dx\,\left(
  {\partial S\over\partial\phi(x)}-j(x)\right)\Delta\phi(x)-
  {1\over2}\int\!dx\,dy\,\Delta\phi(x)N(x,y)\Delta\phi(y)\right)
  \label{ZAD}
\end{eqnarray}
The integration over ${\cal D}\Delta\phi$ is Gaussian and can be
easily taken.  The result reads,
\begin{equation}
  G(x_1\ldots x_n)=Z_0^{-1}\int\!{\cal D}\phi\,
  \phi(x_1)\ldots\phi(x_n)
  \exp\left(-{1\over2}
  \left({\delta S\over\delta\phi}-j\right)\cdot N^{-1}\cdot
  \left({\delta S\over\delta\phi}-j\right)\right)
  \label{G_int}
\end{equation}
The normalization factor $Z_0$, analogously, can be written in the form,
\begin{equation}
  Z_0=\int\!{\cal D}\phi\,\exp\left(-{1\over2}
  \left({\delta S\over\delta\phi}-j\right)\cdot N^{-1}\cdot
  \left({\delta S\over\delta\phi}-j\right)\right)
  \label{Z0_int}
\end{equation}
Eqs.\ (\ref{G_int},\ref{Z0_int}) are the our final result and can be
interpreted as follows.  The Green function $G(x_1\ldots x_n)$ of the
full quantum system can be computed by by taking average over an
ensemble of classical field configurations.  Each field
configuration $\phi(x)$ in the statistical ensemble has the following
weight, 
\[
  \rho[\phi(x)]=Z_0^{-1}\exp\left(-{1\over2}
  \left({\delta S\over\delta\phi}-j\right)\cdot N^{-1}\cdot
  \left({\delta S\over\delta\phi}-j\right)\right)
\]
Therefore, we have reduced the problem of computing the Green
functions in the full quantum theory to the problem of averaging over
a classical statistical ensemble with a known distribution of field
configurations.  In literature \cite{CalzettaHu,GreinerMuller}, Eqs.\ 
(\ref{G_int},\ref{Z0_int}) are usually interpreted in term of a
stochastic equation.  In this language, the dynamics of the field
$\phi$ is described by a Langevin equation with a noise term,
\begin{equation}
  {\delta S\over\delta\phi}-j=\xi
  \label{Langevin_gen}
\end{equation}
where the stochastic source $\xi$ is Gaussian distributed and is
characterized by the following correlation function,
\begin{equation}
  \langle\xi(x)\xi(y)\rangle=N(x,y)
  \label{noise_gen}
\end{equation}
One should note, however, that $N(x,y)$, in general, depends on
$\phi$, and the Langevin equation where the noise kernel depends on
the solution is not very well-defined: to generate the noise, one need
to know the solution beforehand, while the latter cannot be known
without having generated the noise.  The proper understanding of Eqs.\ 
(\ref{Langevin_gen},\ref{noise_gen}) should be always Eqs.\ 
(\ref{G_int},\ref{Z0_int}).  In further discussion, we will use the
stochastic language of Eqs.\ (\ref{Langevin_gen},\ref{noise_gen}),
implicitly assuming that their meaning is given by Eqs.\ 
(\ref{G_int},\ref{Z0_int}).

The physical interpretation of $j$ and $N$ can be seen from Eqs.\ 
(\ref{Langevin_gen},\ref{noise_gen}).  First, according to Eq.\ 
(\ref{noise_gen}), $N$ characterizes the amplitude of the noise coming
from the $\chi$ degrees of freedom that acts on $\phi$.  To counteract
this noise, $j$ must contain a damping that dissipates the energy
coming from the noise.  We will hence call $j$ and $N$ the damping
term and the noise kernel, respectively.

The Langevin dynamics found above will serve as the starting point of
our discussion of the hot gauge theory case.  Let us recall here that
the condition $\Delta\phi\ll\phi$ is the basic assumption in our
derivation of Eqs.\ (\ref{G_int}) and (\ref{Z0_int}), so one should
explicitly verify that this condition is satisfied in the case of hot
gauge theories.

\section{Soft modes in hot gauge theories}
\label{sec:core}

Let us apply this formalism developed in the previous section to the
case hot gauge theories.  The Lagrangian of the theory is,
\[
  {\cal L}=-{1\over4}F^a_{\mu\nu}F^{a\mu\nu}={1\over2g^2}
  \text{tr}\,F_{\mu\nu}F^{\mu\nu}
\]
where, beside the component notation $A^a_\mu$, we use the standard
matrix notation,
\begin{eqnarray}
  A_\mu & = & -igA^a_\mu t^a,\qquad F_{\mu\nu}=-igF^a_{\mu\nu}t^a
  \nonumber \\
  F_{\mu\nu} & = & \partial_\mu A_\nu+\partial_\nu A_\mu+[A_\mu,A_\nu]
  \label{FA}
\end{eqnarray}
where $t^a$ are the generators of the gauge group $G$ in fundamental
representation.

Consider the theory at some temperature $T$.  We will divide the
system into two, a ``soft'' and a ``hard'', subsystems, and we want
the soft subsystem to contain all non-perturbative modes.  In Refs.\ 
\cite{ASY,HuetSon} we have found that these modes have spatial
momentum of order $g^2T$ and frequency of order $g^4T$.  We also want
most important hard modes, those with spatial momentum and frequency
of order $T$, to be in the hard subsystem.  Therefore, we take an
intermediate momentum scale $p_0=c_1g^2T$, and an intermediate
frequency scale $\omega_0=c_2g^4T$, where $c_1$ and $c_2$ are two
arbitrary large numbers $c_1$, $c_2\gg1$ (which are still
parametrically of order 1), and decompose the field into two parts,
\begin{equation}
  A_\mu\to A_\mu+a_\mu
  \label{Amu_decomp}
\end{equation}
where the new $A_\mu$ is the soft component which contains only modes
with spatial momentum $p<p_0$ and frequency $\omega<\omega_0$, and the
hard $a_\mu$ contains other modes.  Note that the decomposition
(\ref{Amu_decomp}) does not respect gauge invariance, therefore one
has to check the gauge invariance of the effective soft dynamics after it
is derived.

The field tensor is decomposed as
\[
  F_{\mu\nu}\to F_{\mu\nu}+f_{\mu\nu}+[a_\mu,a_\nu]
\]
where the new $F$ is related to the new $A$ as in (\ref{FA}),
$f_{\mu\nu}=D_\mu a_\nu-D_\nu a_\mu$, where $D_\mu a_\nu$ is the
covariant derivative of the hard field $a$ on the background $A$,
\[
  D_\mu a_\nu=\partial_\mu a_\nu+[A_\mu,a_\nu]\equiv
  (\partial_\mu+{\cal A}_\mu)a_\nu
\]  
in which ${\cal A}_\mu$ is in the adjoint representation with the
matrix elements ${\cal A}_\mu^{ab}=-gf^{abc}A_\mu^c$.  In general, we
will use calligraphic letters for matrices in adjoint representation.

To derive correctly the soft dynamics, it is crucial to know the
typical magnitude of the field $A_\mu$ and the soft field tensor
$F_{\mu\nu}$.  These quantity can been estimated by using the
fluctuation-dissipation theorem \cite{ASY,HuetSon}.  We quote here the
only the result.  By order of magnitude, spatial components $A^a_i\sim
gT$ (or, in matrix notations, $A_i\sim g^2T$).  This guarantees $A$ to
be non-perturbative: in the expression for the field tensor,
$F_{ij}=\partial_iA_j-\partial_jA_i+[A_i,A_j]$, the last quadratic
term is as important as the first two linear terms.  The component
$A_0$ is usually set to 0 by gauge fixing, however we will relax this
condition by only demanding $A_0$ to be of order $g^4T$, i.e.\ by a
factor of $g^2$ smaller than the spatial components.  This allows one
to perform the gauge transformation on $A$, $A_\mu\to UA_\mu
U^{-1}+U\partial_\mu U^{-1}$, with $U$ varying on the same spatial and
time scales as $A$, i.e.\ $(g^2T)^{-1}$ and $(g^4T)^{-1}$ respectively,
without breaking the conditions $A_i\sim g^2T$, $A_0\sim g^4T$.  The
effective theory for $A_\mu$ should be invariant under these ``soft''
gauge transformations.

For the field tensor, the magnetic and electric components have
different orders of magnitude.  The magnetic components are
$F_{ij}\sim \partial_iA_j+A_i^2\sim g^4T^2$, while the electric ones
are: $F_{0i}=\partial_0A_i-\partial_iA_0+[A_0,A_i]\sim g^6T^2$, where
we have made use of the fact that $\partial_0\sim A_0\sim g^4T$.

Now let us apply the influence functional technique described in Sec.\ 
\ref{sec:review}, identifying the soft field $A_\mu$ with $\phi$ and
the hard field $a_\mu$ with $\chi$.  To integrate over $a_\mu$, one
divides the action into the soft and the hard parts,
\[
  S=S_{\text{s}}[A_\mu]+S_{\text{h}}[a_\mu,A_\mu]
\]
where $S_{\text{s}}$ contains only $A$,
\[
  S_{\text{s}}[A_\mu]={1\over2g^2}\text{tr}\int\!dx\,F^2_{\mu\nu}
\]
while $S_{\text{h}}$ contains the action for $a$ and the interaction
between $A$ and $a$,
\[
  S_{\text{h}}[a_\mu,A_\mu]={1\over2g^2}\text{tr}\int\!dx\,\left(
      f^2_{\mu\nu}+2F_{\mu\nu}[a^\mu, a^\nu]+\cdots\right)
\]
where dots denotes terms cubic or of higher orders on $a$.  Following
the general formalism, one should compute the influence functional,
\[
  \text{e}^{S_{\text{IF}}[A_\mu]}=\int\limits_C\!{\cal D}a_\mu\,
  \exp(iS_{\text{h}}[a_\mu,A_\mu])
\]
To be able to integrate over $a_\mu$, we have to fix the gauge for the
hard field first.  At finite temperatures, it is
most convenient to work in the background Coulomb gauge, where $a$
satisfies the constraint $D_ia^i=0$, since in this gauge only physical
degrees of freedom (transverse gauge bosons) are propagating.  One
writes
\begin{equation}
  \text{e}^{S_{\text{IF}}[A_\mu]}=\int\limits_C\!{\cal D}a_\mu\,
  \exp\Bigl(iS_{\text{h}}[a_\mu,A_\mu]+
  \log\det({\bf D}^2+[a^i,D_i]\Bigr)\delta({\cal D}_ia^i)
  \label{logdet}
\end{equation}
where the commutator $[a^i,D_i]$ is taken with respect to group
indices.  The determinant is usually rewritten into the form of an
integration over the Faddeev-Popov ghost, but there is no need to
introduce the latter in our context.

Let us first show that $\log\det({\bf D}^2+[a^i,D_i])$ is irrelevant.
For this end we note that since the operator staying in the
determinant does not contain any time derivative, it is equal to the
integration of the 3d Euclidean $\log\det$ over the contour,
\[
  \log\det({\bf D}^2+[a^i,D_i])=\int\limits_C\!dt\,
  \text{log}\,\text{det}^{\text{3d}}({\bf D}^2+[a^i,D_i])
\]
The 3d Euclidean determinant can be computed by calculating the
one-loop Feynman diagrams in the 3d theory, which is a quite
complicated task.  Fortunately, for our purpose we need only an
order-of-magnitude estimation, which is
\[
  \log\det({\bf D}^2+a^iD_i)\sim\int\!d{\bf x}\,\Bigl(
  g^2F^a_{ij}F^a_{ij}+g^2f^a_{ij}f^a_{ij}\Bigr)
\]
The second term, $g^2f^a_{ij}f^a_{ij}$, is small compared to
$S_{\text{h}}[a,A]$ in Eq.\ (\ref{logdet}) which contains
$f^a_{ij}f^a_{ij}$.  The first term enters $S_{\text{IF}}[A]$ but is
suppressed compared to $S_{\text{s}}[A]$.  Therefore, the determinant
can be safely ignored.  This result can be interpreted in the
following way: the determinant is the ghost loop, but in the Coulomb
gauge the ghost is not a propagating degree of freedom, therefore its
contribution is the same as at zero temperature, which is suppressed
by a power of the coupling constant.

The influence functional, thus, can be written as
\begin{equation}
  \text{e}^{iS_{\text{IF}}[A_\mu]}=\int\!{\cal D}a_\mu\,
  \text{e}^{iS_{\text{h}}[a_\mu,A_\mu]}\delta({\cal D}_ia^i)
  =\int\!{\cal D}a_\mu\,{\cal D}\lambda\,
  \exp\left(iS_{\text{h}}[a_\mu,A_\mu]+
  \lambda^a{\cal D}_ia^{ai}\right)
  \label{SIF_int}
\end{equation}
where we introduced an auxiliary variable $\lambda^a$.  According to
Eqs.\ (\ref{j}) and (\ref{N}), to compute $j_\mu$ and $N_{\mu\nu}$ one
need to take first and second derivatives of $S_{\text{IF}}$ over
$A_\mu$.

\subsection{Computation of the damping term $j$}

Let us find $j^{a\nu}=\delta S[A]/\delta A^a_\mu$.  One writes,
\[
  j^{a\nu}=-{1\over2Z_0}\int\!{\cal D}a_\mu\,{\cal D}\lambda\,
  \left({\delta\over\delta A^a_{1\nu}}+
  {\delta\over\delta A^a_{2\nu}}\right)
  \left(S_{\text{h}}[a_\mu,A_\mu]+
  i\lambda^a{\cal D}_ia^{ai}\right)
  \exp\Bigl(iS_{\text{h}}[a_\mu,A_\mu]+i\lambda^a{\cal D}_ia^{ai}\Bigr)
\]
\begin{equation}
  ={1\over2Z_0}\int\!{\cal D}a_\mu\,{\cal D}\lambda\,
  \Bigl(D_\mu[a^\mu,a^\nu]^a+[a_\mu,f^{\mu\nu}]^a+
  [a^i,\lambda]^a\delta_{i\nu}\Bigr)\Big|_{1+2}
  \exp\Bigl(iS_{\text{h}}[a_\mu,A_\mu]+i\lambda^a{\cal D}_ia^{ai}\Bigr)
  \label{jnu}
\end{equation}
where, for the simplicity of notations, we denote the sum over the
upper and lower parts of the contour by the index ``1+2''.  We will be
interested primarily in the spatial components of $j^\nu$.  Running
ahead, it turns out that the contributions from the upper and the
lower parts of the contour are equal to each other, so one needs to
compute only one of them and then multiply the result by a factor of
2.  Let us concentrate on the contribution from the upper part of the
contour (corresponding to the index 1).  Using the constraint
$D_ia^i=0$ imposed by the integration over $\lambda$, one can rewrite
Eq.\ (\ref{jnu}) into the form,
\begin{eqnarray}
  j^i & = & Z_0^{-1}\int\!{\cal D}a_\mu\,{\cal D}\lambda\,
  \left(2D_\mu[a^\mu,a^i]-[D_0a^0,a^i]+[a^0,D_ia^0]-[a^j,D_ia^j]
  +[a^i,\lambda]\right)\times \nonumber \\
  & & \times\exp\left(iS_{\text{h}}[a,A]+i\lambda^aD_ia^{ai}\right)
  \label{j_commut}
\end{eqnarray}
Let us introduce the propagators on the soft background,
\[
  G_{ab}^{\mu\nu}(x,y)=\langle a^{a\mu}(x)a^{b\nu}(y)\rangle=
  Z_0^{-1}\int\limits_C\!{\cal D}a_\mu\,a^{a\mu}(x)a^{b_\nu}(y)
  \exp\left(iS_{\text{h}}[a,A]+i\lambda^aD_ia^{ai}\right)
\]
\[
  H_{ab}^i(x,y)=\langle a^{ai}(x)\lambda^b(y)\rangle=
  Z_0^{-1}\int\limits_C\!{\cal D}a_\mu\,a^{ai}(x)\lambda^b(y)
  \exp\left(iS_{\text{h}}[a,A]+i\lambda^aD_ia^{ai}\right)
\]
(for avoiding cumbersome notations, we use lower group indices for $G$
and $H$).  In subsequent equations only the 11 component of these
propagators will enter, so for the simplicity we will not write the
index ``11'' explicitly.  From Eq.\ (\ref{j_commut}), one sees that
$j^i$ can be expressed via these propagators as
\begin{equation}
  j^{ai}(x)=gf^{abc}\biggl[2D_\mu^x G_{bc}^{\mu i}(x,x)+
  H_{bc}^i(x,x)+
  \Bigl(D^x_iG_{bc}^{jj}(x,y)-D^x_iG_{bc}^{00}(x,y)-
  D^x_0G^{0i}_{bc}(x,y)\Bigr)\Big|_{y\to x}\biggr]
  \label{j_prop}
\end{equation}
where we define the covariant derivatives of $G(x,y)$ with respect to
$x$ and $y$ as follows,
\[
  D^x_\mu G_{ab}(x,y)={\partial\over\partial x^\mu}G_{ab}(x,y)+
  {\cal A}^{ac}(x)G_{cb}(x,y)
\]
\[
  D^y_\mu G_{ab}(x,y)={\partial\over\partial y^\mu}G_{ab}(x,y)-
  G_{ac}(x,y){\cal A}^{cb}(y)
\]
In order to find $j$, thus, one need to find the propagators $G$ and
$H$ on the background field $A$.  Let us specify the accuracy one
needs to know these propagators.  Since the Langevin equation has the
form $\delta S_{\text{s}}[A]/\delta A^a=j^a+\xi^a$, we expect $j$ to
be of the same order as $\delta S/\delta A^a\sim D^2A^a\sim g^5$,
therefore, one needs to find $G$ and $H$ up to contributions of order
$g^4$.

In Appendix \ref{app:propagator} we showed that the
temperature-dependent parts of $G^{00}$, $G^{0i}$, and $H$ are of
order $g^6$ and thus can be ignored in Eq.\ (\ref{j_prop}).  Also,
while $G(x,x)$ is of order 1, its $x$-dependent part is of
proportional to $g^4$ and varies on the spatial scale $(g^2T)^{-1}$,
so $D_iG(x,x)\sim g^6$.  Therefore, Eq.\ (\ref{j_prop}) can be
simplified to
\begin{equation}
  j^{ai}(x)=gf^{abc}D^x_i G^{jj}_{bc}(x,y)|_{y\to x}
  \label{jxy}
\end{equation}
Instead of $x$ and $y$, it is convenient to introduce the new
coordinates $X$ and $s$,
\[
  x=X+{s\over2},\qquad y=X-{s\over2}
\]
For simplicity, we will denote the propagator in these coordinates as
$G(X,s)$.  As a function of $s$, $G$ varies on the scale of $T^{-1}$,
while as a function of $X$ the variation is over the spatial scale
$(g^2T)^{-1}$ and time scale $(g^4T)^{-1}$.  In the coordinates $X$
and $s$, Eq.\ (\ref{jxy}) reads,
\begin{eqnarray}
  j^{ai}(X) & = & gf^{abc}\left({\partial\over\partial s^i}+
  {1\over2}{\partial\over\partial X^i}+
  {\cal A}_i\left(X+{s\over2}\right)\right)G^{jj}_{bc}(X,s)= \nonumber \\
  & = & gf^{abc}\left({\partial\over\partial s^i}+{\cal A}_i(X)\right)
  G^{jj}_{bc}(X,s)\Bigg|_{s\to0}
  \label{jXs}
\end{eqnarray}
To derive the last expression, we again have made use of the fact that
$\partial_XG\sim g^6$ when $s\to0$ ($x\to y$) and can be neglected.
Thus, to compute $j$, one need to find the spatial components of $G$
up to the order $g^4$ and plug into Eq.\ (\ref{jXs}).

In Appendix \ref{app:propagator} the propagator $G(X,s)$ is computed
with the accuracy we need, i.e.\ up to terms of order $g^4$
inclusively.  The result is somehow complicated, but some parts of
it allow intuitive physical interpretation.  First, there is a
contribution to $G$ that simply comes from gauge invariance.  To
extract this contribution from $G$, one writes,
\begin{equation}
  G_{ab}^{ij}(x,y)=U^{ab}(x,y)G_0^{ij}(x,y)+\tilde{G}^{ij}_{ab}(x,y)
  \label{Gsplit}
\end{equation}
where $U^{ab}(x,y)$ are the matrix elements of the Wilson line
connecting two point $x$ and $y$ in the adjoint representation,
\[
  U^{ab}(x,y)=T\exp\left(-\int\limits_y^x\!dz^\mu\,
  {\cal A}_\mu(z)\right)
\]
$G_0$ is the propagator in the absence of the background field (i.e.,
when $A=0$), and $\tilde{G}$ is of order $g^4$.  All corrections of
order $g^2$, thus, is due to the Wilson line $U^{ab}(x,y)$, while at
the same time there is a part of the $g^4$ corrections that cannot be
attributed to the latter.  In particular, when $x=y$, or $s=0$,
propagator differs from $G_0$ (which is $X$-independent) by an amount
of order $g^4$.  Substituting Eq.\ (\ref{Gsplit}) into Eq.\ 
(\ref{jXs}), one sees that only the $\tilde{G}$ term makes
contribution to $j$.  Since $\tilde{G}$ is already of order $g^4$, to
the order of $g$ that we are interested in one can write,
\[
  j^{ai}(X)=gf^{abc}{\partial\over\partial s^i}
  \tilde{G}_{bc}^{jj}(X,s)\bigg|_{s\to0}
\]
It is convenient to perform the Wigner transformation on $\tilde{G}$,
\[
  \tilde{G}(X,p)=\int\!ds\,\text{e}^{ips}\tilde{G}(X,s)
\]
so $j$ can be written in the form,
\[
  j^{ai}(X)=2igf^{abc}\int\!{d^4p\over(2\pi)^4}\,p^i
  \tilde{G}_{bc}^{jj}(X,p)
\]
Now let us quote the result of Appendix \ref{app:propagator} for
$\tilde{G}$ (more precisely, on its 11 component)
\begin{eqnarray}
  \tilde{G}^{ij}_{ab}(X,p)i & = & -if^{abc}\delta n_{\bf p}^c(X)
  \left(\delta^{ij}-{p^ip^j\over{\bf p}^2}\right)
  2\pi\delta(p_0^2-{\bf p}^2)-{\cal F}_{ij}(X)\left(
  {1\over|{\bf p}|}{\partial G_s\over\partial{\bf p}}
  -{G_s\over2{\bf p}^2}\right)+ \nonumber \\
  & & +{p^k\over{\bf p}^2}
  \left({\cal F}_{ik}(X)p^j-{\cal F}_{jk}(X)p^i\right)
  \left({1\over2|{\bf p}|}{\partial G_s\over\partial|{\bf p}|}
  -{G_s\over{\bf p}^2}\right)
  \label{propagator}
\end{eqnarray}
where $\delta n^a_{\bf p}(X)$, with $X=(t,{\bf x})$, is defined as
follows,
\begin{equation}
  \delta n^a_{\bf p}(t,{\bf x})=
  -g{\partial n_{\bf p}\over\partial|{\bf p}|}
  \int\limits_0^\infty\!du\,U^{ab}(t,{\bf x};t-u,{\bf x}-u{\bf v})
   {\bf v}\cdot{\bf E}^b(t-u,{\bf x}-u{\bf v})
  \label{deltan_nonlocal}
\end{equation}
and $G_s$ is defined in Eqs. (\ref{Gs}) and depends on ${\bf p}$ only
via its module $|{\bf p}|$.

First let us substitute Eq.\ (\ref{propagator}) into the formula for
$j^i$.  One notices immediately that for computing $\tilde{G}^{jj}$
only the first term in the RHS of Eq.\ (\ref{propagator}) matters:
other terms are odd under the permutation $i\leftrightarrow j$.  One
finds,
\begin{equation}
  j^{ai}(X)=2C_2(G)g\int\!{d{\bf p}\over(2\pi)^3}\,
  v^i\delta n_{\bf p}^a(X)
  \label{j=vn}
\end{equation}
where $C_2(G)$ is defined by $f^{acd}f^{bcd}=C_2(G)\delta^{ab}$ (for
the SU($N$) group, $C_2(\text{SU}(N))=N$).  Eq.\ (\ref{j=vn}) allows
us to interpret $\delta n$ as the deviation of the distribution
function from thermal equilibrium.  Using Eq.\ 
(\ref{deltan_nonlocal}), $j$ can be rewritten via the soft field in
the following way,
\[
  j^{ai}(t, {\bf x})=-2C_2(G)g^2\int\!{d{\bf p}\over(2\pi)^3}
  \int\limits_0^\infty\!du\,
  {\partial n_{\bf p}\over\partial|{\bf p}|}
  U^{ab}(t,{\bf x};t-u,{\bf x}-u{\bf v})
  v^iv^jE^{bj}(t-u,{\bf x}-u{\bf v})
\]
One can now write $d{\bf p}={\bf p}^2\,d|{\bf p}|\,d{\bf v}$, where
${\bf v}$ is the velocity that corresponds to ${\bf p}$, ${\bf v}={\bf
  p}/|{\bf p}|$, and take the integration over $d|{\bf p}|$.  The
result reads,
\[
  j^{ai}(t,{\bf x})={C_A(G)g^2T^2\over12\pi}\int\!du\,d{\bf v}\,
  U^{ab}(t,{\bf x};t-u,{\bf x}-u{\bf v})
  v^iv^jE^{bj}(t-u,{\bf x}-u{\bf v})
\]
Now in the integral it is convenient change the variables $u$ and
${\bf v}$ to a new variable ${\bf y}={\bf x}-u{\bf v}$, and one finds,
\begin{equation}
  j^{ai}(t,{\bf x})={C_2(G)g^2T^2\over12\pi}\int\!d{\bf y}\,
  U^{ab}(t,{\bf x};t-|{\bf x}-{\bf y}|, {\bf y})
  {(x-y)^i(x-y)^j\over|{\bf x}-{\bf y}|^4}
  E^{bj}(t-|{\bf x}-{\bf y}|, {\bf y})
  \label{j_nonlocal}
\end{equation}
It is natural to assume that the integral in Eq.\ 
(\ref{deltan_nonlocal}) is saturated by such ${\bf y}$ that $|{\bf
  x}-{\bf y}|\sim(g^2T)^{-1}$.  In fact, for larger distances between
${\bf y}$ and ${\bf x}$ the Wilson line becomes a rapidly varying
function, so the contribution from theses distances vanishes.  Since
$|{\bf x}-{\bf y}|\sim(g^2T)^{-1}$, one can replace, in Eq.\ 
(\ref{j_nonlocal}), $t-|{\bf x}-{\bf y}|$ by $t$, as the typical time
scale of the evolution of soft modes is much larger than
$(g^2T)^{-1}$.  Therefore, the dependence of $j^a$ on the soft field
can be considered as local in time (but still nonlocal in space),
\[
  j^{ai}(t,{\bf x})={C_2(G)g^2T^2\over12\pi}\int\!d{\bf y}\,
  U^{ab}(t,{\bf x};t,{\bf y}){(x-y)^i(x-y)^j\over|{\bf x}-{\bf y}|^4}
  E^{bj}(t, {\bf y})
\]
One can substitute $E^i=-\dot{A}^i-D_i A_0$ to this equation.  The
term with $A_0$ can be taken by integration by part and the result
vanishes due to the identity
\begin{equation}
  D^x_i\left(U^{ab}(t,{\bf x};t,{\bf y})
  {(x-y)^i(x-y)^j\over|{\bf x}-{\bf y}|^4}\right)=0
  \label{gamma_id}
\end{equation}
so one obtains finally,
\begin{equation}
  j^{ai}(t,{\bf x})=-{C_2(G)g^2T^2\over12\pi}\int\!d{\bf y}\,
  U^{ab}(t,{\bf x};t,{\bf y}){(x-y)^i(x-y)^j\over|{\bf x}-{\bf y}|^4}
  \dot{A}^{bj}(t, {\bf y})
  \label{j_final}
\end{equation}

Note that we have computed $j$ by multiplying the ``1'' component in
Eq.\ (\ref{jnu}) by 2.  One can check that the ``2'' component, which
can be found by the same method as described above, is equal to the
``1'' component.

\subsection{Computation of the noise kernel $N$}

The noise kernel, according to the general formula (\ref{N}), is equal
to
\begin{equation}
  N^{ij}_{ab}(x,y)=-{i\over4}
  {\delta^2\over\delta A_i^a(x)\delta A_j^b(y)}S_{\text{IF}}[A]
  \bigg|_{11+12+21+22}
  \label{Ngauge}
\end{equation}
where we have indicated that the sum over all possible locations of
$x$ and $y$ (on upper or lower parts of the contour) should be taken.
Similar to the case of $j$, it turns out that all four terms
corresponding to $x$ and $y$ on different parts of the contour $C$ are
equal, so one need to compute only one of them and then multiply the
result by 4.  Let us pick the 12 component and write,
\[
  N^{ij}_{ab}(x,y)=-i{\delta^2\over\delta A^a_{1i}(x)
  \delta A^b_{2j}(y)}S_{\text{IF}}[A]
\]
In the following we will not write the indices 1 and 2, implicitly
assume that $x$ is on the upper and $y$ is on the lower parts of the
contour.  Using Eq.\ (\ref{SIF_int}), one finds,
\begin{eqnarray}
  N^{ij}_{ab}(x,y) & = &
  \langle[a^k(x),D_ia^k(x)]^a[a^l(y),D_ja^l(y)]^b\rangle-
  \langle[a^k(x),D_ia^k(x)]^a\rangle\langle[a^l(y),D_ja^l(y)]^b\rangle=
  \nonumber \\
  & = & f^{acd}f^{bef}\left(
  G^{kl}_{ce}(x,y)\partial^x_i\partial^y_jG^{kl}_{df}(x,y)+
  \partial^y_jG^{kl}_{cf}(x,y)\partial^x_iG^{kl}_{de}(x,y)\right)
  \label{NGG}
\end{eqnarray}
where we have replaced the covariant derivative $D$ by the simple
derivative $\partial$.  It is possible to make such replacement since
we will be interested only in the leading order.  In fact,
$D=\partial+{\cal A}$, and for hard field like $a$, $\partial\sim T$
while ${\cal A}\sim g^2T$.

Since we are interested in the case when $x-y\sim(g^2T)^{-1}$, we need
to know the propagator in the regime of such separation between $x$
and $y$.  To the leading order on $g$, the calculation of the
propagator for $x-y\sim(g^2T)^{-1}$ is carried out in Appendix
\ref{app:propagator2}.  The result, for the 12 component, reads
\begin{eqnarray*}
  G^{ij}_{ab}(x,y) & =
  \int\!{d{\bf p}\over(2\pi)^32|{\bf p}|}\,
  \Bigl(f_{\bf p}(x)f_{\bf p}^{-1}(y)\Bigr)^{ab}
  \Bigl(\delta^{ij}-{p^ip^j\over{\bf p}^2}\Bigr)
  \Bigl( & n_{\bf p}\text{e}^{-i|{\bf p}|(t-t')+
  i{\bf p}({\bf x}-{\bf y})}+ \nonumber \\
  & & +(1+n_{\bf p})\text{e}^{i|{\bf p}|(t-t')-
  i{\bf p}({\bf x}-{\bf y})}\Bigr)
\end{eqnarray*}
where $x=(t,{\bf x})$, $y=(t',{\bf y})$, and $f_{\bf p}(x)$ is, for
each value of the coordinate $x$, a group element in adjoint
representation which is defined as follows,
\begin{equation}
  f^{ab}_{\bf p}(x)=U^{ab}(t,{\bf x}; 0,{\bf x}-{\bf v}t)
  \label{U_def}
\end{equation}
(as before, ${\bf v}={\bf p}/|{\bf p}|$) provided the background $A$
is turned of for $t<0$.  Now let us substitute this propagator to Eq.\ 
(\ref{NGG}).  The first term in $N$ is
\[
  f^{acd}f^{bef}G^{kl}_{ce}(x,y)
  \partial^x_i\partial^y_jG^{kl}_{df}(x,y)
  =f^{acd}f^{bef}
  \int\!{d{\bf p}d{\bf p}'\over(2\pi)^64|{\bf p}||{\bf p}'|}\,p^ip'^j
  \left(\delta^{ij}-{p^ip^j\over{\bf p}^2}\right)
  \left(\delta^{ij}-{p'^ip'^j\over{\bf p}'^2}\right)\times
\]
\[
  \times\left(f_{\bf p}(x)f_{\bf p}^{-1}(y)\right)^{ce}
  \left(f_{\bf p'}(x)f_{\bf p'}^{-1}(y)\right)^{df}
  \left(n_{\bf p}\text{e}^{-i|{\bf p}|(t-t')+
  i{\bf p}({\bf x}-{\bf y})}+
  (1+n_{\bf p})\text{e}^{i|{\bf p}|(t-t')-
  i{\bf p}({\bf x}-{\bf y})}\right)\times
\]
\begin{equation}
  \times\left(n_{{\bf p}'}\text{e}^{-i|{\bf p}'|(t-t')+
  i{\bf p}'({\bf x}-{\bf y})}+
  (1+n_{{\bf p}'})\text{e}^{i|{\bf p}'|(t-t')-
  i{\bf p}({\bf x}-{\bf y})}\right)
  \label{Nintpp'}
\end{equation}
Notice that ${\bf p}$, ${\bf p}'$ are of order $T$, while $t-t'$ and
$x-y$ are of order $(g^2T)^{-1}$, therefore the integrand in Eq.\ 
(\ref{Nintpp'}) is a rapidly varying function unless ${\bf p}$ and
${\bf p}'$ are close to each other.  The integral, thus, is dominated
by this region.  Denoting ${\bf p}-{\bf p'}=\Delta{\bf p}$, and notice
that when $\Delta{\bf p}$ is small one has $|{\bf p}|-|{\bf p}'|={\bf
  v}\cdot\Delta{\bf p}$, one can write Eq.\ (\ref{Nintpp'}) into the
following form,
\begin{eqnarray*}
  f^{abc}f^{bef}G^{kl}_{ce}(x,y)
  \partial^x_i\partial^y_jG^{kl}_{df}(x,y) & = &
  4f^{acd}f^{bef}\int\!
  {d{\bf p}\,d\Delta{\bf p}\over(2\pi)^64{\bf p}^2}
  \Bigl(f_{\bf p}(x)f_{\bf p}^{-1}(y)\Bigr)^{ce}
  \Bigl(f_{\bf p}(x)f_{\bf p}^{-1}(y)\Bigr)^{df}\times \\
  & & \times n_{\bf p}(1+n_{\bf p})p^ip^j
  \text{e}^{-i{\bf v}\cdot\Delta{\bf p}(t-t')+
  i\Delta{\bf p}\cdot({\bf x}-{\bf y})}
\end{eqnarray*}
Now one can take the integration over $\Delta{\bf p}$, which yields a
delta function,
\begin{eqnarray*}
  f^{abc}f^{bef}G^{kl}_{ce}(x,y)
  \partial^x_i\partial^y_jG^{kl}_{df}(x,y) & = &
  4f^{acd}f^{bef}\int\!{d{\bf p}\over(2\pi)^34{\bf p}^2}
  \Bigl(f_{\bf p}(x)f_{\bf p}^{-1}(y)\Bigr)^{ce}
  \Bigl(f_{\bf p}(x)f_{\bf p}^{-1}(y)\Bigr)^{df}\times \\
  & & \times n_{\bf p}(1+n_{\bf p})p^ip^j
  \delta({\bf x}-{\bf y}-{\bf v}(t-t'))
\end{eqnarray*}
Due to the the delta function, the integrand is non-zero only when
$x-y=(t-t',{\bf x}-{\bf y})$ is proportional to $v=(1,{\bf v})$.  In
this case, one can verify using Eq.\ (\ref{U_def}) the following
identity,
\[
  f_{\bf p}(x)f_{\bf p}^{-1}(y) = U(x,y)
\]
which is independent of ${\bf p}$ and can be taken out of the
integration.  Therefore, 
\begin{eqnarray*}
  f^{acd}f^{bef}G^{kl}_{ce}(x,y)
  \partial^x_i\partial^y_jG^{kl}_{df}(x,y) & = &
  f^{acd}f^{bef}U^{ce}(x,y)U^{df}(x,y)\times \\
  & & \times\int\!{d{\bf p}\over(2\pi)^3}\,v^iv^jn_{\bf p}(1+n_{\bf p})
  \delta({\bf x}-{\bf y}-{\bf v}(t-t'))
\end{eqnarray*}
The pre-integral factor can be written as
$-\text{tr}\,T^aU(x,y)T^bU^{-1}(x,y)$ where $T^a$ are the generators
of the gauge group in adjoint representation, $(T^a)^{bc}=-f^{abc}$.
To write this factor in a simple way, one notice that the following
identity holds for any matrix $U$ in the adjoint representation of the
group $G$,
\[
  \text{tr}\,T^aUT^bU^{-1}=-C_2(G)U^{ab}
\]
The second term in Eq.\ (\ref{NGG}) can be computed analogously, and
the result is equal to that of the first term.  One obtains,
\[
  N^{ab}_{ij}(t,{\bf x};t',{\bf y})=2U^{ab}(x,y)
  \int\!{d{\bf p}\over(2\pi)^3}v^iv^jn_{\bf p}(1+n_{\bf p})
  \delta({\bf x}-{\bf y}-{\bf v}(t-t'))
\]
Taking the integration over $d{\bf p}$, one finds,
\begin{equation}
  N_{ab}^{ij}={C_2(G)g^2T^3\over12\pi}
  U^{ab}(t,{\bf x}; t', {\bf y})
  {(x-y)^i(x-y)^j\over|{\bf x}-{\bf y}|^4}
  \Bigl(\delta(t-t'-|{\bf x}-{\bf y}|)+
  \delta(t-t'+|{\bf x}-{\bf y}|)\Bigr)
  \label{N_nonlocal}
\end{equation}
Now since ${\bf x}-{\bf y}\sim(g^2T)^{-1}$ which is much smaller than
the time scale interested in, one can ignore the time non-locality in
$N$ and replace Eq.\ (\ref{N_nonlocal}) with a local one,
\begin{equation}
  N_{ab}^{ij}(t,{\bf x}; t',{\bf y})={C_2(G)g^2T^3\over6\pi}
  U^{ab}(t,{\bf x}; t,{\bf y})
  {(x-y)^i(x-y)^j\over|{\bf x}-{\bf y}|^4}
  \delta(t-t')
  \label{N_final}
\end{equation}
Using the same method, it is easy to check that the 11, 21 and 22
components in Eq.\ (\ref{Ngauge}) is equal to the 12 component, so
$N$ can be computed by taking the 12 component and multiplying to 4, as
we have done.  Therefore, Eq.\ (\ref{N_final}) is the noise kernel we
want to find.

\subsection{Verification of the initial assumptions $\Delta A\ll A$}

To make sure that we have derived the right equation, one has to
verify that the basic assumption used in the derivation, namely, that
the difference between the values of the soft field $A$ in the upper
and lower parts of the contour is small, $\Delta A\ll A$, is valid
(more precisely, that the path integral is saturated by field
configurations with $\Delta A\ll A$).  For this end, let us recall
that the integral over $\Delta A$ is taken with the weight,
\[
  \exp\left(-{1\over2}\int\!dx\,dy\,
  \Delta A^a(x)N_{ab}(x,y)\Delta A^b(y)\right)
\]
So, knowing the order of magnitude of $N$, one can estimate how large
is a typical $\Delta A$.  As we have found before, $N(x,y)$ is
nonzero when $|{\bf x}-{\bf y}|\sim x_0-y_0\sim (g^2T)^{-1}$, and for
these values of $x$ and $y$ the typical value of $N$ is, according to
Eq.\ (\ref{N_nonlocal}),
\begin{equation}
  N(x,y)\sim g^2T^3\cdot{1\over(x-y)^2}\cdot{1\over(x_0-y_0)}
  \sim g^8T^6
  \label{N_typ}
\end{equation}
Now to find the typical value of $\Delta A$, one writes
\[
  \int\!dx\,dy\,N(\Delta A^a)^2\sim1
\]
provided the integral is limited to a region of $x$ and $y$ with
linear size of order $(g^2T)^{-1}$.  Given the typical value of $N$ in
Eq.\ (\ref{N_typ}), one obtains the estimate $\Delta A^a\sim g^4T$.
Recall that the typical value of the soft field is $A^a\sim gT$, one
sees that $\Delta A\ll A$.

Now we will explain why we have treated the terms linear and quadratic
on $\Delta A$ on equal footing while neglecting cubic and higher terms
in the expansion of $S_{\text{IF}}$.  Naively, if $\Delta A\ll A$, the
linear term $j\cdot\Delta A$ is much larger than the quadratic term
$\Delta A\cdot N\cdot\Delta A$.  However, we have seen that there is a
delicate cancelation in $j$: the leading and next-to-leading orders
gives no contributions to $j$, the first nonzero contribution comes
from the order $g^4$.  In contrast, there is no such cancelation in
$N$.  More careful order-of-magnitude analysis show that these two
terms are equally important.  Since there is no cancelation in the
second term $\Delta A\cdot N\cdot\Delta A$, higher terms are certainly
more suppressed than the second-order one and can be neglected.

\subsection{Fixing the ambiguity of the Langevin equation: the 
  Fokker-Planck equation}

Above we have computed $j$ and $N$, which are the two quantities one
need to know in order to write down the Langevin equation for the soft
modes.  The Langevin equation has the following form,
\begin{equation}
  D_kF^a_{ki}(t,{\bf x})=
  -{C_2(G)g^2T^2\over12\pi}\int\!
  d{\bf y}{(x-y)^i(x-y)^j\over|{\bf x}-{\bf y}|^4}
  U^{ab}(t,{\bf x};t,{\bf y})\dot{A}^{bj}(t,{\bf y})
  +\xi^a(t,{\bf x})
  \label{Lang_naive1}
\end{equation}
\begin{equation}
  \langle\xi^{ai}(t,{\bf x})\xi^{bj}(t',{\bf y})\rangle=
  {C_2(G)g^2T^3\over6\pi}
  {(x-y)^i(x-y)^j\over|{\bf x}-{\bf y}|^4}
  U^{ab}(t,{\bf x};t,{\bf y})\delta(t-t')
  \label{Lang_naive}
\end{equation}
Notice that in the LHS of Eq.\ (\ref{Lang_naive1}) should stays
$\delta S[A]/\delta A^i=-D_0F_{0i}+D_kF_{ki}$, but $D_0F_{0i}$ is
smaller than $D_kF_{ki}$ by a factor of $g^4$, so the former can be
neglected.  These equations have been first written down in Ref.\ 
\cite{HuetSon}.  However, an important point has been missed there:
the Langevin equation with local noise and local damping, where the
noise kernel is a function of the coordinate is ill-defined, in the
sense that different discretization of this Langevin equation leads to
different evolution in the continuum limit.  Therefore, one has to be
more careful in interpreting Eqs.\ 
(\ref{Lang_naive1},\ref{Lang_naive}) as the equation describing soft
dynamics in hot gauge plasma.

The ambiguity of the Langevin equation with local noise and damping
depending on the coordinate is known in literature \cite{Risken}, but
since the topic is important for our discussion, let us illustrate it
on a simple example of Brownian motion of a one dimensional particle
describing by the equation,
\begin{equation}
  \gamma(x)\dot{x}+V'(x)=\xi
  \label{Lang_1dmain}
\end{equation}
Note there is no $\ddot{x}$ term as in Eq.\ (\ref{Lang_naive1}), which
means that we are in the regime of over-damping.  The coefficient
$\gamma$ represents friction and varies with $x$.  The noise kernel
will be chosen to be local in time, but also $x$-dependent,
\begin{equation}
  \langle\xi(t)\xi(t')\rangle=2T\gamma(x)\delta(t-t')
  \label{Lang_1dsource}
\end{equation}
The unique way to characterize the Brownian motion is via the
Fokker-Planck equation, which describes the diffusion of the particle.
Denoting $\rho(t,x)$ to be distribution function of particle (the
probability of finding the particle at the point $x$ at time moment
$t$), the Fokker-Planck equation is the one describing the evolution
of $\rho(t,x)$ with $t$.

In Appendix \ref{app:Langevin_asym} we demonstrate the ambiguity of
the Fokker-Planck equation for this simple one-dimensional case.  We
check that for two different ways of discretizing Eqs.
(\ref{Lang_1dmain},\ref{Lang_1dsource}), one obtains two different
Fokker-Planck equation.  For example, if one approximates Eqs.
(\ref{Lang_1dmain},\ref{Lang_1dsource}) by the following
finite-difference equations
\[
  \gamma(x(t)){x(t+\Delta t)-x(t)\over\Delta t}=\xi(t)
\]
\[
  \langle\xi(t)\xi(t+n\Delta t)\rangle=
  {2T\gamma(x(t))\over\Delta t}\delta_{n0}
\]
and then takes the limit $\Delta t\to0$, then the Fokker-Planck
equation is
\begin{equation}
  {\partial\rho(t,x)\over\partial t}=T{\partial^2\over\partial x^2}
  \left({\rho(t,x)\over\gamma(x)}\right)+
  {\partial\over\partial x}\left({V'(x)\over\gamma(x)}\rho(t,x)\right)
  \label{FP1d1}
\end{equation}
while, if one chooses to compute $\gamma$ at $x(t+\Delta t)$ instead of
$x(t)$, i.e.\ use the following discretization,
\[
  \gamma(x(t+\Delta t)){x(t+\Delta t)-x(t)\over\Delta t}=\xi(t)
\]
\[
  \langle\xi(t)\xi(t+n\Delta t)\rangle=
  {2T\gamma(x(t+\Delta t))\over\Delta t}\delta_{n0}
\]
then the Fokker-Planck equation becomes
\begin{equation}
  {\partial\rho\over\partial t}=T{\partial\over\partial x}
  \left({1\over\gamma^3}{\partial\over\partial x}\Bigl(
  \gamma^2\rho\Bigr)\right)+
  {\partial\over\partial x}\left({V'\over\gamma}\rho\right)
  \label{FP1d2}
\end{equation}
It is clear that this equation is different from Eq.\ (\ref{FP1d1})
which has been derived using another scheme of approximation.  In
general, by choosing different discretization scheme, one could end
up with the equation
\[
  {\partial\rho\over\partial t}=T{\partial\over\partial x}\biggl(
  \gamma^{\alpha-1}{\partial\over\partial x}\Bigl(
  \gamma^{-\alpha}\rho\Bigr)\biggr)+
  {\partial\over\partial x}\left({V'\over\gamma}\rho\right)
\]
with arbitrary value of $\alpha$.  Eq.\ (\ref{FP1d1}) corresponds to
$\alpha=1$ while Eq.\ (\ref{FP1d2}) corresponds to $\alpha=-2$.  Which
value of $\alpha$ one should take for the Fokker-Planck equation in
the case of gauge theory?

We argue that the value $\alpha=0$ is the correct one.  Note that both
discretizations of Eqs. (\ref{Lang_1dmain},\ref{Lang_1dsource}) are
asymmetric: the damping and the source kernel are evaluated at either
end of the time interval $(t,t+\Delta t)$.  It seems that the more
``correct'' choice would be the one where these quantities are
evaluated at the middle of this interval.  In Appendix
\ref{app:Langevin} we consider such a scheme.  In this scheme, we add
a small second derivative term $\epsilon\ddot{x}$ to the Langevin
equation and discretize it in the following way,
\[
  \epsilon{x(t+\Delta t)-2x(t)+x(t-\Delta t)\over(\Delta t)^2}+
  \gamma(x(t)){x(t+\Delta t)-x(t-\Delta t)\over2\Delta t}+
  V'(x(t))=\xi(t)
\]
\[
  \langle\xi(t)\xi(t+n\Delta t)\rangle=2T\gamma(x(t))\delta_{n0}
\]
The small second-derivative term must be added for the prescription to
have a limit at $\Delta t\to0$.  Moreover, in the case of gauge theory
the second time derivative of $A$ does really present in the equation
(we ignore it in Eq.\ (\ref{Lang_naive1}) since it is much smaller than
spatial derivatives).  The limit $\Delta t\to0$ should be taken before
the limit $\epsilon\to0$.  In Appendix \ref{app:Langevin} we show that
the Fokker-Planck equation, for this prescription, has the form,
\begin{equation}
  {\partial\rho(t,x)\over\partial t}=
  {\partial\over\partial x}\left({T\over\gamma(x)}
  {\partial\over\partial x}\rho(t,x)\right)+
  {\partial\over\partial x}\left({V'(x)\over\gamma(x)}\rho(t,x)\right)
  \label{Lang_1d}
\end{equation}
which can be written as
\[
  {\partial\rho\over\partial t}+{\partial j\over\partial x}=0
\]
where $j=-{T\over\gamma}{\partial\rho\over\partial x}-
{V'\over\gamma}\rho$ is the probability current, which is a sum of the
current coming from diffusion (the first term) and from the overall
drift due to the force $V'$ (the second term).

The static solution to Eq.\ (\ref{Lang_1d}) is given by the barometric
formula,
\[
  \rho(t,x)\sim\text{e}^{-{V(x)\over T}}
\]
One can check this by verifying that for this probability
distribution, the probability current $j$ vanishes.

Eqs.\ (\ref{Lang_naive1},\ref{Lang_naive}) is the multi-dimensional
version of Eq.\ (\ref{Lang_1dmain},\ref{Lang_1dsource}) with a
particular choice of $\gamma$.  It is straightforward to generalize
Eq.\ (\ref{Lang_1d}) and write the Fokker-Planck equation for our
case.  Let $\rho(t,{\bf A})$ be the distribution function, then
\begin{equation}
  {\partial\over\partial t}\rho(t,{\bf A})=
  \int\!d{\bf x}\,d{\bf y}\,{\delta\over\delta A^{ai}({\bf x})}
  \left((\gamma^{-1})^{ab}_{ij}({\bf x}, {\bf y};{\bf A})\left(
  T{\delta\rho(t,{\bf A})\over\delta A^{bj}({\bf y})}+
  {\delta V({\bf A})\over\delta A^{bj}({\bf y})}
  \rho(t,{\bf A})\right)\right)
  \label{FP_cont}
\end{equation}
where $\gamma^{-1}$ is the inverse of the operator $\gamma$,
\[
  \gamma^{ab}_{ij}({\bf x},{\bf y};{\bf A})={C_2(G)g^2T^2\over12\pi}
  U^{ab}({\bf x},{\bf y};{\bf A})
  {(x-y)^i(x-y)^j\over|{\bf x}-{\bf y}|^4}
\]
and $V({\bf A})={1\over4}\int\!d{\bf x}\,F^a_{ij}F^a_{ij}$ is the
static energy of the field configuration $A$.

The Fokker-Planck equation, Eq.\ (\ref{FP_cont}), is the central
result of this paper.

While the Langevin equation (\ref{Lang_naive1},\ref{Lang_naive}) is
ambiguous, the Fokker-Planck equation (\ref{FP_cont}) defines the
evolution of soft modes in a unique way.  There is still a problem
with the degeneracy of $\gamma$, which is related to the gauge
invariance of the Fokker-Planck equation.  We defer this question to
the end of this section.

Similarly with the one-dimensional case, Eq.\ (\ref{FP_cont})
possesses a static solution,
\begin{equation}
  \rho({\bf A})\sim\exp\left(-{V({\bf A})\over T}\right)
  \label{dim_red}
\end{equation}
Therefore, the Fokker-Planck equation we have derived is consistent
with the thermodynamics of the soft modes: it is known from
dimensional reduction \cite{Shaposhnikov} that the statics of soft
modes is described by a classical ensemble with the barometric
distribution as in Eq.\ (\ref{dim_red}).

We did not present the rigorous proof that Eq.\ (\ref{FP_cont}) is the
correct choice of the Fokker-Planck equation.  To rigorously show the
latter, one need to go one step backward from Eq.\ 
(\ref{Lang_naive1},\ref{Lang_naive}), i.e.\ not neglect the
non-locality in damping and noise kernel and carefully study the
behavior of the solution on time scales much larger than that of the
non-locality.  Unfortunately, this program seems to be very
complicated.  We believe, however, that the arguments presented above
convincingly showed that Eq.\ (\ref{FP_cont}) is the real
Fokker-Planck equation describing the soft modes.

Let us finish this subsection by noticing the Markovian character in
the evolution of the soft modes.  The time evolution of $A$ can be
visualized as a large numbers of transitions $t\to t+\Delta t\to
t+\Delta t_2\to\cdots$ where each transition occurs independently from
others.  We emphasize that the Markovian character is manifest only at
large time scales (such as $(g^4T)^{-1}$): on the time scale of
$(g^2T)^{-1}$ the evolution of soft modes is certainly non-Markovian.
However, there is no substantial evolution during such small time
intervals.

\subsection{Gauge invariance of effective soft dynamics and degeneracy
  of $\gamma$}

In this subsection, we will show that the Fokker-Planck equation we
have found is gauge invariant.  Recall that for $A$ we did not fix the
$A_0=0$ gauge, but only impose a rather relaxed condition that
$A_0\sim g^4$, which allows one to make any soft gauge transformation
with parameter varying on the spatial scale $(g^2T)^{-1}$ and time
scale of order $(g^4T)^{-1}$.  Therefore, the effective dynamics of
the gauge field must not break the invariance under these gauge
transformations.  The verification of this fact is a crucial test for
our formalism, since, as we already noted, the division of the system
into soft and hard modes is not a gauge invariant operation and it is
important to check that the final result is nevertheless gauge
invariant.

Let us first define what is is soft gauge invariance in term of the
distribution function $\rho({\bf A})$.  The gauge invariance of $\rho$
simply means that $\rho({\bf A})$ is the same for equivalent field
configurations ${\bf A}$,
\begin{equation}
  \rho({\bf A})=\rho({\bf A}^U)
  \label{rho_gauge_inv}
\end{equation}
where $A_i=UA_iU^{-1}+U\partial_iU^{-1}$.  We will now check that the
RHS of the Fokker-Planck equation is also gauge-independent, so if
$\rho$ satisfies Eq.\ (\ref{rho_gauge_inv}) at some time moment it
will satisfy at later times.

Under gauge transformation, $\delta/\delta A$ transforms as
\[
  {\delta\over\delta A^U}=
  {\delta\over\delta A}{\delta A\over\delta A^U}=
  {\cal U}^{-1}{\delta\over\delta A}={\delta\over\delta A}{\cal U}
\]
where ${\cal U}$ is the matrix $U$ in the adjoint representation.  The
matrix $\gamma({\bf x},{\bf y}; {\bf A})$ transforms as
\[
  \gamma({\bf x},{\bf y};{\bf A}^U)=
  {\cal U}^{-1}({\bf x})\gamma({\bf x},{\bf y};{\bf A}){\cal U}({\bf y})
\]
so $\gamma^{-1}$ should transform analogously.  One sees that the RHS
of Eq.\ (\ref{FP_cont}) is invariant under gauge transformations.

However, the real situation is more complex.  In Eq.\ (\ref{FP_cont})
we implicitly assume $\gamma$ is not degenerate in writing
$\gamma^{-1}$.  However, it is easy to see that $\gamma$ is
degenerate!  In fact, assume one wants to find
$\beta_i=\gamma^{-1}_{ij}\alpha_j$.  For that one needs to solve the
equation,
\begin{equation}
  \int\!d{\bf y}\,\gamma_{ij}({\bf x},{\bf y};{\bf A})
  \beta_j({\bf y})=\alpha_i({\bf x})
  \label{alphabeta}
\end{equation}
Taking the derivative $D_i$ on this equation and summing over $i$,
making use of the identity $D^x_i\gamma_{ij}({\bf x},{\bf y})=0$ (see
Eq.\ (\ref{gamma_id})), one finds $D_i\alpha_i=0$.  Therefore Eq.\ 
(\ref{alphabeta}) possesses a solution only when $D_i\alpha_i=0$.

Moreover, provided there exists a $\beta$ that satisfies Eq.\ 
(\ref{alphabeta}), one can easily check that
$\beta'_i=\beta_i+D_i\eta$, where $\eta$ is an arbitrary function, is
also a solution.  Therefore, we conclude that $\gamma$ is degenerate:
$\gamma^{-1}$ can be defined only when it acts on a function
$\alpha_i$ satisfying the requirement $D_i\alpha_i=0$, and the result
itself is not uniquely defined but only up to a full covariant
derivative.

Then how one can understand the Fokker-Planck equation
(\ref{FP_cont})?  One will see that this equation is still
well-defined just because of the gauge invariance.

Let us verify that one can act $\gamma^{-1}$ on
${\delta\rho\over\delta A}+{\delta V\over\delta A}\rho$.  One has to
check that
\[
  D_i\left(T{\delta\rho\over\delta A^i}+{\delta V\over\delta A^i}\rho
  \right)=0
\]
It is easy to see that $D_i{\delta V\over\delta A^i}\sim
D_iD_jF_{ij}=0$.  Therefore, what really needs to be checked is
$D_i{\delta\rho\over\delta A^i}=0$.  This condition comes from the
gauge invariance of $\rho$.  In fact, under infinitesimal
transformations, Eq.\ (\ref{rho_gauge_inv}) reads,
\[
  \rho(A_i)=\rho(A_i+D_i\alpha)
\]
for any function $\alpha$.  This is equivalent to
\[
  \int\!d{\bf x}\,{\delta\rho\over\delta A_i({\bf x})}
  D_i\alpha({\bf x})=0
\]
Integrating by part this equation, and making use of the arbitrariness
of $\alpha$, one finds,
\[
  D_i{\delta\rho\over\delta A_i({\bf x})}=0
\]
which is exactly the condition for $\gamma^{-1}{\delta\rho\over\delta
A}$ to exist.  Therefore, we have checked that one can define
\begin{equation}
  f^a_i({\bf x})=(\gamma^{-1})^{ab}_{ij}
  \left(T{\delta\rho\over\delta A^b_j}+
  {\delta V\over\delta A^b_j}\rho\right)
  \label{f_def}
\end{equation}
in the sense that $\gamma f=T{\delta\rho\over\delta A}+{\delta
V\over\delta A}\rho$, if $\rho$ satisfy the gauge invariance condition
(\ref{rho_gauge_inv}).

There is still an ambiguity in the definition of $f^a_i({\bf x})$:
this quantity is defined only up to a full covariant derivative.  One
may ask whether different choice of this quantity lead to different
evolution of $\rho$.  We will see that this is not the case, provided
one defines $\gamma^{-1}$ in a gauge invariant way.  The latter means
the requirement that $f$ transforms covariantly under gauge
transformations,
\[
  f({\bf x};{\bf A}^U)=U({\bf x})f({\bf x};{\bf A})U^{-1}({\bf x})
\]
After imposing this condition, $f$ is still not defined uniquely.  One
can replace
\begin{equation}
  f^a_i({\bf x};{\bf A})\to f^a_i({\bf x};{\bf A})+
  D_ic^a({\bf x};{\bf A})
  \label{freplace}
\end{equation}
where the only requirement on $c({\bf x};{\bf A})$ is that it
satisfies the equation
\begin{equation}
  D^{U}_ic({\bf x};{\bf A}^U) = U({\bf x})D_ic({\bf x};{\bf A})
  U^{-1}({\bf x})
  \label{c_gauge}
\end{equation}
for any function $U({\bf x})$.  After the replacement
(\ref{freplace}), the RHS of the Fokker-Planck equation
(\ref{FP_cont}) picks up a term proportional to
\begin{equation}
  \int\!d{\bf x}\,{\delta\over\delta A^a_i({\bf x})}
  D_ic^a({\bf x};{\bf A})
  \label{deltaDc}
\end{equation}
For the Fokker-Planck equation to be uniquely defined, one needs to
show that this expression vanishes.

Let us consider Eq.\ (\ref{c_gauge}) for infinitesimal gauge
transformations $U=1+\alpha$, $A_i\to A_i-D_i\alpha$.  It reads,
\[
  D_ic({\bf x};{\bf A})-[D_i\alpha({\bf x}),c({\bf x};{\bf A})]-
  D^x_i\int\!d{\bf y}\,
  {\delta c({\bf x};{\bf {\bf A}})\over\delta A_j({\bf y})}
  D_j\alpha({\bf y})=D_ic({\bf x},{\bf A})+
  [\alpha({\bf x}),D_ic({\bf x};{\bf A})]
\]
Taking the integral over $d{\bf y}$ by part, one finds
\[
  D^x_i\int\!d{\bf y}\,D^y_j
  {\delta c({\bf x};{\bf A})\over\delta A_j({\bf y})}\alpha({\bf y})=
  D^x_i[\alpha({\bf x}),c({\bf x};{\bf A})]
\]
which implies,
\[
  \int\!d{\bf y}\,D^y_i
  {\delta c({\bf x};{\bf A})\over\delta A_i({\bf y})}\alpha({\bf y})=
  [\alpha({\bf x}),c({\bf x};{\bf A})]
\]
In components, this equation reads,
\[
  \int\!d{\bf y}\,\left({\partial\over\partial y^i}
  {\delta c^a({\bf x};{\bf A})\over\delta A^b_i({\bf y})}+
  igf^{bcd}A^c_i({\bf y})
  {\delta c^a({\bf x};{\bf A})\over\delta A^d_i({\bf y})}\right)
  \alpha^b({\bf y})=-igf^{abc}\alpha^b({\bf x})c^c({\bf x};A)
\]
Since this equation must be valid for any function $\alpha$, one
finds,
\[
  {\partial\over\partial y^i}
  {\delta c^a({\bf x};{\bf A})\over\delta A^b_i({\bf y})}+
  igf^{bcd}A^c_i({\bf y})
  {\delta c^a({\bf x};{\bf A})\over\delta A^d_i({\bf y})}=
  -igf^{abc}c^c({\bf x})\delta({\bf x}-{\bf y})
\]
Now putting $a=b$ and take the sum over $a$, and taking ${\bf y}\to
{\bf x}$, we obtain,
\[
  {\partial\over\partial y^i}{\delta c^a({\bf x};{\bf A})\over
  \delta A^a_i({\bf y})}\Bigg|_{{\bf y}\to{\bf x}}+
  igf^{acd}A^c_i({\bf x})
  {\delta c^a({\bf x};{\bf A})\over\delta A^d_i({\bf x})}=0
\]
Taking the integral over $x$, and noticing that
\[
  {\partial\over\partial y^i}
  {\delta c^a({\bf x};{\bf A})\over\delta A^a_i({\bf y})}
  \Bigg|_{{\bf y}\to{\bf x}}+
  {\partial\over\partial x^i}
  {\delta c^a({\bf x};{\bf A})\over\delta A^a_i({\bf y})}
  \Bigg|_{{\bf y}\to{\bf x}}
\]
is a full derivative, one finds,
\[
  \int\!d{\bf x}\,\left(
  {\delta\partial_ic^a({\bf x};{\bf A})\over\delta A^a_i({\bf x})}
  +igf^{abc}A^b_i({\bf x})
  {\delta c^c({\bf x},{\bf A})\over\delta A^c_i({\bf x})}\right)=0
\]
The LHS of this equation is the same as the expression we want to show
to be vanished, Eq.\ (\ref{deltaDc}), thus we have demonstrated that
the RHS of the Fokker-Planck equation is independent of the way we
choose $\gamma^{-1}$.  So, despite the degeneracy of $\gamma$, the
Fokker-Planck equation is still well-defined and gauge invariant.

\section{Conclusion}
\label{sec:conclusion}

In this paper we have derived the Fokker-Planck equation describing
the non-perturbative real-time dynamics of the soft modes in hot gauge
plasma, Eq.\ (\ref{FP_cont}).  With some reservations, the same
dynamics can be described by the Langevin equation, Eq.\ 
(\ref{Lang_naive1},\ref{Lang_naive}).  The dynamics is diffusive in
nature, where the word ``diffusive'' refers to the evolution in the
space of field configurations: assuming a given configuration of the
soft field at some time moment, the field will ``diffuse'' and
eventually fill the whole space of field configurations with the
thermal distribution during a time interval of order $(g^4T)^{-1}$.
The behavior is totally different from that predicted by the classical
field equation, in contrast to the case of scalar theory \cite{Aarts}
where there is a time regime where the classical field equation can be
applied.  The difference between the behavior of the gauge field and
of a scalar field originates from the fact that in the gauge case the
soft modes are in the regime of over-damping \cite{ASY} while in the
scalar case they are not.  We believe that the understanding the
diffusive nature of the soft field evolution in gauge theories is
crucial for any discussion of physical processes involving real-time
dynamics of soft modes, for example, baryon number violation at high
temperatures.

While from Fokker-Planck equation is the fundamental equation of
long-distance real-time dynamics at high temperatures, from the point
of view of numerical simulations the Langevin equation
(\ref{Lang_naive1},\ref{Lang_naive}) may have its advantage.  However,
even the latter is hardly suitable for real simulations due to the
spatial non-locality.  Moreover, the Langevin equation is ambiguous.
At this moment, it seems that the only practical way is to simulate a
classical theory whose soft modes have the same dynamics as the hot
quantum theory.  Some proposals have been put forward in this
direction.  One is to simply simulate a classical lattice theory of a
gauge field \cite{Arnold}, another is to simulate the kinetic of a
plasma consisting of a soft field and a collection of hard particles
\cite{HuMuller}.  The latter approach might be more economical, though
the rigorous proof that the dynamics of the soft modes in this picture
can be made identical to that of the quantum theory is still lacking.
The Fokker-Planck equation derived in this paper can serve as a
criterion for choosing the right simulation procedure, which,
hopefully, will leads to the ultimate solution to the problem of hot
baryon number violation rate.

\acknowledgments

The author thanks P.~Arnold, A.~Krasnitz, G.D.~Moore, and I.~Smit for
useful discussions.  This work was supported by the U.S.\ Department
of Energy, Grant No.\ DE-FG03-96ER40956.

\appendix

\section{Hard propagators on soft background: small separations}
\label{app:propagator}

In this Appendix, we will compute the propagators of the hard field
$a$ and the auxiliary field $\lambda$ on the soft background $A$,
$G\sim\langle aa\rangle$ and $H\sim\langle a\lambda\rangle$.  We will
need to find these propagators up to the correction of order $g^4$
inclusively.  Since $A\sim g^2T$, this means that one need to find
also correction of the second order with respect to $A$.

We will be especially interested in spatial components of $G$,
\[
  G_{ab}^{ij}=Z_0^{-1}\int\!{\cal D}a\,{\cal D}\lambda\,
  a^{ai}(x)a^{bj}(y)
  \exp\Bigl(iS_{\text h}[a,A]+i\lambda^aD_ia^{ai}\Bigr)
\]
We will first compute this propagator at tree level, i.e.\ neglecting
the non-linear (on $a$) terms in $S_{\text{h}}$, and then show that
the result does not change when one turns on the interaction between
hard modes.

\subsection{Tree level}

The convenient way to compute the propagators at tree level is to
introduce a source $j$ coupled linearly to $a$ and study the linear
response of the system in the presence of this source.  In this way,
the propagators can be computed using the following formulas,
\[
  G^{\mu\nu}_{ab}(x,y)=\langle a^{a\mu}(x)a^{b\nu}(y)\rangle=
  i{\delta\langle a^{a\mu}(x)\rangle_j\over j^b_\nu(y)}
\]
\[
  H^\mu_{ab}(x,y)=\langle a^{a\mu}(x)\lambda^b(y)\rangle=
  i{\delta\langle\lambda^b(y)\rangle_j\over j^a_\mu(x)}
\]
where $\langle\cdots\rangle_j$ denotes the mean value of in the
presence of the source $j$,
\begin{eqnarray}
  \langle a^{a\mu}\rangle_j & = & Z_0^{-1}[j]
  \int\!{\cal D}a^\mu\,{\cal D}\lambda\,a^{a\mu}(x)
  \exp\left(iS_{\text{h}}[a,A]+i\lambda^aD_ia^{ai}-
   ij^a_\mu a^{a\mu}\right) \nonumber \\
  \langle\lambda^a\rangle_j &= &Z_0^{-1}[j]
  \int\!{\cal D}a^\mu\,{\cal D}\lambda\,\lambda^a(x)
  \exp\left(iS_{\text{h}}[a,A]+i\lambda^aD_ia^{ai}-
  ij^a_\mu a^{a\mu}\right)
  \label{mean_a}
\end{eqnarray}
where
\[
  Z_0[j]=\int\!{\cal D}a^\mu\,{\cal D}\lambda\,
  \exp\left(iS_{\text{h}}[a,A]+i\lambda^aD_ia^{ai}-
  ij^a_\mu a^{a\mu}\right)
\]
At tree level, finding these main values is equivalent to finding the
saddle point of the exponent in Eq.\ (\ref{mean_a}), i.e.\ the solution
to the following set of equation,
\begin{eqnarray}
  & & D^2a^\mu-D^\mu(D\cdot a)+2{\cal F}^{\mu\nu}a_\nu-
  \delta^{\mu i}D_i\lambda-j^\mu=0 \nonumber \\
  & & D_ia^i=0
  \label{field_eq_source}
\end{eqnarray}
where $D^2=D_\mu D^\mu$.  We will need to know the solution to Eqs.
(\ref{field_eq_source}) to the accuracy of $g^6$ (we will see that
this accuracy is necessary for finding the propagators with the
accuracy of $g^4$).  First let us find $a^0$. Substituting $\mu=0$ to
Eq.\ (\ref{field_eq_source}), one obtains,
\[
  -{\bf D}^2a^0+2{\cal F}_{0i}a^i=j^0
\]
where ${\cal F}$ denotes the field tensor in adjoint representation,
${\cal F}^{ab}=-gf^{abc}F^c$, from which one can  express $a^0$ via
other variables,
\begin{equation}
  a^0=-{\bf D}^{-2}j^0+2{\bf D}^{-2}({\cal F}_{0i}a^i)
  \label{a0}
\end{equation}
Notice that $a^i$ is hard, and ${\cal F}_{0i}\sim g^6$, so without
making error of order $O(g^6)$, this equation can be written as
\begin{equation}
  a^0=-{\bf D}^{-2}j^0+2{\cal F}_{0i}\vec{\partial}^{-2}a^i
  \label{a0_sol}
\end{equation}
Since $G^{00}\sim\partial a^0/\partial j^0$, and ${\cal F}_{0i}\sim
g^6$, one finds immediately
\[
  G^{00}=-i{\bf D}^{-2}+O(g^6)=G^{00}|_{T=0}+O(g^6)
\]
where we have made use of the fact that the leading term in $G^{00}$
does not contain time derivative and thus is temperature-independent.
From Eq.\ (\ref{a0_sol}) one can also estimate $G^{0i}$.  Indeed,
$G^{0i}\sim\partial a^0/\partial j^i$, and from Eq.\ (\ref{a0_sol})
$a^0$ depends on $j^i$ only through $a^i$.  Since $F_{0i}\sim g^6$,
one finds,
\[
 G^{0i}=O(g^6)
\]

Let us turn to the equation for $a^i$.  One writes,
\[
  D^2a^i+D_i(D_0a^0)-2{\cal F}_{ij}a^j+D_i\lambda=j^i
\]
Substituting $a^0$ found in Eq.\ (\ref{a0}), one obtains
\begin{equation}
  D^2a^i+D_i(-D_0{\bf D}^{-2}j^0+
  2{\cal F}_{0j}\partial_0\vec{\partial}^{-2}a^j)
  -2{\cal F}_{ij}a^j-D_i\lambda=j^i
  \label{eq4ai}
\end{equation}
Let us express $\lambda$ via the other unknowns.  For this end we
apply $D_i$ to Eq.\ (\ref{eq4ai}) and make use of the constraint
$D_ia^i=0$.  One finds,
\begin{equation}
  -({\cal D}_i{\cal F}_{ij})a^j-{\bf D}^2D_0{\bf D}^{-2}j^0
  -{\bf D}^2\lambda=D_ij^i
  \label{lambda_pre}
\end{equation}
Denoting ${\cal D}_i{\cal F}_{ij}={\cal J}_j$, which is of order $g^6$,
one can extract $\lambda$ from Eq.\ (\ref{lambda_pre}),
\[
  \lambda=-D_0{\bf D}^{-2}j^0-
  {\bf D}^{-2}(D_ij^i)-{\cal J}^i\vec{\partial}^{-2}a^i
\]
From this one also finds
\[
  H_{ab}^i(x,y)=H_{ab}^i(x,y)|_{T=0}+O(g^6)
\]
i.e.\ the temperature-dependent part of $H_i$ is of order $g^6$.
Substitute $a^0$ and $\lambda$ to the equation for $a^i$, the latter
reads,
\begin{equation}
  D^2a_i+2{\cal F}_{0k}\partial_0\vec{\partial}^{-2}
  \partial_ia_k-2{\cal F}_{ik}a^k+
  {\cal J}_k\vec{\partial}^{-2}\partial_ia^k=
  j^i-D_i{\bf D}^{-2}D_kj^k+O(g^8)
  \label{D2ai}
\end{equation}
To find the Green function, now we adapt the technique for finding the
kinetic equation (see \cite{LifshitzPitaevskii} to the case with
background field.  First, according to Eq.\ (\ref{D2ai}), the spatial
components of the Green function $G$ satisfies the following equations
\[
  \left({\partial\over\partial x^\lambda}+{\cal A}_\lambda(x)\right)^2
  G_{ij}(x,y)+
  \left(2{\cal F}_{0k}(x){\partial\over\partial x^0}+
  {\cal J}_k(x)\right)
  \left({\partial^2\over\partial{\bf x}^2}\right)^{-1}
  {\partial\over\partial x^i}G_{kj}(x,y)-
\]
\[
  -2{\cal F}_{ik}(x)G_{kj}(x,y)=-i\delta^{\text{tr}}_{ij}(x,y)+O(g^8)
\]
The same equation can be also written as
\[
  G_{ij}(x,y)\left({\loarrow{\partial}\over\partial y^\lambda}-
  {\cal A}_\lambda(y)\right)^2
  -G_{ik}(x,y){\loarrow{\partial}\over\partial y^j}
  \left({\loarrow{\partial}^2\over\partial{\bf y}^2}\right)^{-1}
  \left({\loarrow{\partial}\over\partial y^0}
  2{\cal F}_{0k}(y)+{\cal J}_k(y)\right)-
\]
\[
  -2G_{ik}(x,y){\cal F}_{kj}(y)=-i\delta^{\text{tr}}_{ij}(x,y)+O(g^8)
\]
where $\delta^{\text{tr}}_{ij}(x,y)=\delta_{ij}\delta(x,y)-D_i{\bf
  D}^{-2}D_j\delta(x,y)$.  For simplicity we did not write the contour
indices (1 and 2) in our equations, with the implicit understanding
that one can put these indices back in at the end of the calculations.
Let us subtract the two equations from each others, and divide the
result by 2.  The RHS vanishes while there are 3 different
contributions in the LHS which we will denote as
\begin{equation}
  I_1={1\over2}
  \left({\partial\over\partial x^\lambda}+{\cal A}_\lambda(x)\right)^2
  G_{ij}(x,y)-{1\over2}G_{ij}(x,y)\left(
  {\loarrow{\partial}\over\partial y^\lambda}-
  {\cal A}_\lambda(y)\right)^2
  \label{I1}
\end{equation}
\begin{eqnarray}
  I_2 & = & \left({\cal F}_{0k}(x){\partial\over\partial x^0}+
  {1\over2}{\cal J}_k(x)\right)
  \left({\partial^2\over\partial{\bf x}^2}\right)^{-1}
  {\partial\over\partial x^i}G_{kj}(x,y)+ \nonumber \\ 
  & & +G_{ik}(x,y){\loarrow{\partial}\over\partial y^j}
  \left({\loarrow{\partial}^2\over\partial{\bf y}^2}\right)^{-1}
  \left({\loarrow{\partial}\over\partial y^0}{\cal F}_{0k}(y)+
  {1\over2}{\cal J}_k(y)\right)
  \label{I2}
\end{eqnarray}
\begin{equation}
  I_3=-\Bigl(F_{ik}(x)G_{kj}(x,y)-G_{ik}(x,y)F_{kj}(y)\Bigr)
  \label{I3}
\end{equation}
The obtained equation reads,
\begin{equation}
  I_1+I_2+I_3 = 0
  \label{collisionless}
\end{equation}

First, let us consider $I_1$.  It is convenient to use, instead of $x$
and $y$, the new coordinates $X$ and $s$ which are related to the old
coordinates as
\[
  x=X+{s\over2},\qquad y=X-{s\over2}
\]
For simplicity we will denote the propagator $G(x,y)$ in the new
coordinates as $G(X,s)$.  As a function of $s$, $G(X,s)$ varies on the
scale of $T^{-1}$, but $G$ is a slowly varying function of $X$ which
changes on the spatial scale of $(g^2T)^{-1}$ and time scale of
$(g^4T)^{-1}$.  We will evaluate each contribution $I_1$, $I_2$ and
$I_3$ to the accuracy of $O(g^6)$.  First, consider $I_1$.  Expanding
Eq. (\ref{I1}), making use of the fact that ${\cal A}\sim g^2T$,
$\partial_X\sim g^2T$, $\partial_s\sim T$, one finds,
\begin{eqnarray}
  I_1 & = & {\partial\over\partial X^\mu}{\partial G\over\partial s_\mu}
  +\left[{\cal A}_\mu,{\partial G\over\partial s_\mu}\right]+
  {1\over2}\left\{{\cal A}_\mu,{\partial G\over\partial X_\mu}\right\}+
  {1\over2}\left\{s^\nu\partial_\nu{\cal A}_\mu,
  {\partial G\over\partial s_\mu}\right\}+
  {1\over2}\{s^\mu{\cal A}_\mu,G\}+
  \nonumber \\
  & & + {1\over2}\left[{\cal A}^2,G\right]+
  {1\over2}\{{\cal A}^\mu,s^\nu\partial_\nu{\cal A}_\mu\}G
  \label{I1_pretilde}
\end{eqnarray}
where we $[\cdots]$ denotes commutator,  $\{\cdots\}$ denotes
anti-commutator, and all $A$'s are computed at the point $X$ (from now,
when $A$ and $F$ are written without arguments we will implicitly
assume that they are computed at the point $X$).

To the zeroth order on $g$, the propagator is equal to that in the
absence of the background field, $G_{ab}=G_0\delta^{ab}$, where $G_0$
is defined in Eqs.\ (\ref{G_photon_ij}) and (\ref{G_photon_00}).  When
the background is present, $G_0(x,y)$ is not a gauge invariant
quantity.  To correct this, one can include a Wilson line $U(x,y)$ to
$G_0(x,y)$, and the result will be gauge invariant.  However, there
may be contributions to $G$ which cannot be attributed to the Wilson
line, so we write,
\begin{equation}
  G_{ab}(x,y)=G_0(x,y)U^{ab}(x,y)+\tilde{G}_{ab}(x,y)
  \label{G=G0U}
\end{equation}
Expanding the first term in the RHS of Eq.\ (\ref{G=G0U}) on $g^2$,
one obtains,
\begin{equation}
  G=G_0-(s^\mu{\cal A}_\mu)G_0+{1\over2}(s^\mu{\cal A}_\mu)^2G_0+
  \tilde{G}+O(g^6)
  \label{ansatz_QCD}
\end{equation}
So far, we did not make any assumption on $\tilde{G}$.  All we know
about $\tilde{G}$ is that it is a correction to $G_0$, thus it has
order of magnitude of $g^2$ or less.  We will assume, and verify
subsequently, that $\tilde{G}$ in reality has the order of magnitude
of $g^4$.  Therefore, the $g^2$ correction in $G$ comes entirely from
the Wilson line, while the same is not true for the $g^4$ correction.

Substituting Eq.\ (\ref{ansatz_QCD}) to Eq.\ (\ref{I1_pretilde}) and
neglecting any contributions of order $g^8$ or higher, one finds,
\[
  I_1={\cal D}_\mu{\partial\tilde{G}\over\partial s_\mu}-
  s^\nu{\cal F}_{\mu\nu}{\partial G_0\over\partial s_\mu}+
  {1\over2}s^\nu\{{\cal F}_{\mu\nu},s^\lambda{\cal A}_\lambda\}
  {\partial G_0\over\partial s_\mu}
\]
where ${\cal D}_\mu=\partial/\partial X^\mu+[{\cal A}_\mu(X),\ldots]$.
Now will will split $I_1$ into two parts and evaluate each separately.
The first part is
\[
  I_{1a}={\cal D}_\mu{\partial\tilde{G}\over\partial s_\mu}-
  s^\nu{\cal F}_{\mu\nu}{\partial G_0\over\partial s_\mu}
\]
while the second one is
\begin{equation}
  I_{1b}={1\over2}s^\nu\{{\cal F}_{\mu\nu},s^\lambda {\cal A}_\lambda\}
  {\partial G_0\over\partial s_\mu}
  \label{I1b}
\end{equation}
It is useful to rewrite $I_{1a}$ in the Wigner representation, i.e.\ in
the momentum representation with respect to $s$.  Introducing the
notation,
\[
  G(X,p)=\int\!d^4s\,\text{e}^{ips}G(X,s)
\]
and similar formulas for other $G_0$ and $\tilde{G}$, one finds
\[
  I_{1a}(X,p)=-ip^\mu D_\mu\tilde{G}(X,p)+p^\mu{\cal F}_{\mu\nu}(X)
  {\partial G_0(p)\over\partial p_\nu}
\]
Now, let us take into account the fact that the structure of $G_0$
with respect to group and Lorentz indices is
\begin{equation}
  (G_0)^{ij}_{ab}(p)=\left(\delta^{ij}-{p^ip^j\over{\bf p}^2}\right)
  \delta^{ab}G_s(p)
  \label{Gij}
\end{equation}
where $G_s(p)$ is the propagator of the free massless scalar field.
After some calculations, $I_{1a}$ can be expressed via $G_s$ as
follows,
\begin{equation}
  I_{1a}=\left(\delta^{ij}-{p^ip^j\over{\bf p}^2}\right)
  p^\mu{\cal F}_{\mu\nu}{\partial G_s\over\partial p_\nu}+
  {p^0\over{\bf p}^2}({\cal F}_{0i}p^j+{\cal F}_{0j}p^i)G_s+
  {p^k\over{\bf p}^2}({\cal F}_{ki}p^j+{\cal F}_{kj}p^i)G_s
  \label{1a}
\end{equation}

Now let us turn to $I_{1b}$.  From Eq.\ (\ref{I1b}) one sees that to
compute $I_{1b}$ to the order of $g^6$, one need to consider only the
case when $\mu$, $\nu$ and $\lambda$ are all spatial indices.  It is
more convenient to leave $I_{1b}$ in the coordinate representation,
where
\[
  I_{1b}=-{1\over2}s^l\{{\cal F}_{kl},s^mA_m\}
  {\partial G_0^{ij}\over\partial s^k}
\]
One can substitute $G_0$ to this equation.  For our purpose, it is
sufficient to notice that in the coordinate representation, the most
general tensor form of $G_0$ is
\[
  G_0^{ij}(s)=G_1\delta^{ij}+G_2s^is^j
\]
where $G^1$ and $G^2$ are functions depending only on the module of
${\bf s}$.  In term of these functions, $I_{1b}$ can be rewritten as
\[
  I_{1b}=
  -{1\over2}s^ks^l\left(\{F_{ik},A_l\}s^j+\{F_{jk},A_l\}s^i\right)
  G_2
\]

So, we have computed $I_1$.  For $I_2$, substituting Eq.\ (\ref{Gij}) to
Eq.\ (\ref{I2}), we find, in the momentum representation,
\begin{equation}
  I_2=-{p_0\over {\bf p}^2}({\cal F}_{0i}p^j+{\cal F}_{0j}p^i)G_s-
  {i\over2{\bf p}^2}(p^i{\cal J}_j-p^j{\cal J}_i)G_s
  \label{I2_found}
\end{equation}
Notice that the first term in the RHS of Eq.\ (\ref{I2_found}) cancel
the second term in Eq.\ (\ref{1a}).

Now consider $I_3$.  Making use of Eq.\ (\ref{ansatz_QCD}), $I_3$ can
be divided into two parts, $I_3=I_{3a}+I_{3b}$ where
\[
  I_{3a}=-[{\cal F},G_0]
\]
\[
  I_{3b}=-{1\over2}\left\{s^k\partial_k{\cal F},G_0\right\}+
  \left[{\cal F},(s^k{\cal A}_k)G_0\right]
\]
where the matrix multiplication is understood as occurring for both
(spatial) Lorentz and group indices and both ${\cal A}$ and ${\cal F}$
are computed at the point $X$.  $I_{3a}$ is of order $g^4$, while
$I_{3b}$ contains $O(g^6)$ contributions.  For $I_{3a}$, the
calculations is quite simple.  Using Eq.\ (\ref{Gij}), one obtains,
\[
  I_{3a}=-{\cal F}_{ik}\left(\delta^{kj}-
  {p^kp^j\over{\bf p}^2}\right)G_s+
  \left(\delta^{ik}-{p^ip^k\over{\bf p}^2}\right){\cal F}_{kj}G_s=
  {p^k\over{\bf p}^2}\left({\cal F}_{ik}p^j-{\cal F}_{kj}p^i\right)G_s
\]
One can notice that $I_{3a}$ cancels the last term of $I_{1a}$ in Eq.\ 
(\ref{1a}).  $I_{3b}$ is harder to find.  Substitution of Eq.\ 
(\ref{Gij}) yields
\begin{eqnarray}
  I_{3b} & = & -{1\over2}
  \left(s^l\partial_l{\cal F}_{ik}G_0^{kj}+
  s^lG_0^{ik}\partial_l{\cal F}_{kj}\right)+
  {\cal F}_{ik}s^l{\cal A}_lG_0^{kj}-
  G_0^{ik}s^l{\cal A}_l{\cal F}_{kj} \nonumber \\
  & = & -{1\over2}s^l\Bigl(
  (\partial_l{\cal F}_{ik}-2{\cal F}_{ik}{\cal A}_l)G_0^{kj}+
  G_0^{ik}(\partial_l{\cal F}_{kj}+2{\cal A}_l{\cal F}_{kj})\Bigr)
  \nonumber \\
  & = & -{1\over2}s^l\Bigl({\cal D}_l{\cal F}_{ik}G_0^{kj}+
   G_0^{ik}{\cal D}_l{\cal F}_{kj}-
  \left\{{\cal F}_{ik},{\cal A}_l\right\}G_0^{kj}-
  G_0^{ik}\left\{{\cal A}_l,{\cal F}_{kj}\right\}\Bigr)
  \label{3b}
\end{eqnarray}
We will split $I_{3b}$ further into two parts,
$I_{3b}=I_{3ba}+I_{3bb}$, where
\[
  I_{3ba}=-{1\over2}s^l\left({\cal D}_l{\cal F}_{ik}G_0^{kj}+
  G_0^{ik}{\cal D}_l{\cal F}_{kj}\right)
\]
which, in the momentum representation, reads,
\[
  I_{3aa}=-{i\over2}\left(
  {\cal D}_l{\cal F}_{ik}{\partial\over\partial p^l}G^0_{kj}+
  {\cal D}_l{\cal F}_{kj}{\partial\over\partial p^l}G^0_{ik}\right)=
\]
\begin{eqnarray*}
  & = & -i{\cal D}_k{\cal F}_{ij}{\partial G_s\over\partial p^k}+
  {ip^k\over2{\bf p}^2}\Bigl({\cal D}_l{\cal F}_{ik}p^j+
  {\cal D}_l{\cal F}_{kj}p^i\Bigr){\partial G_s\over\partial p^l}+
  {i\over2{\bf p}^2}\left(p^i{\cal J}_j-p^j{\cal J}_i+
  p^k{\cal D}_k{\cal F}_{ij}\right)G_s- \nonumber \\
  & & -{i\over{\bf p}^4}p^kp^l\left(p^j{\cal D}_l{\cal F}_{ik}+
  p^i{\cal D}_l{\cal F}_{kj}\right)G_s
\end{eqnarray*}
The rest of $I_{3b}$ is denoted as $I_{3ab}$ (which is the terms in
Eq.\ (\ref{3b}) that contains $\{{\cal F},{\cal A}\}$) and can be
most conveniently written in the coordinate representation,
\[
  I_{3ab}={1\over2}s^l\left(
  \left\{{\cal F}_{ik},{\cal A}_l\right\}G_0^{kj}-
  G_0^{ik}\left\{{\cal A}_l,{\cal F}_{kj}\right\}\right)=
  {1\over2}s^ks^l\left(\left\{{\cal F}_{ik},{\cal A}_l\right\}s^j+
  \left\{{\cal F}_{jk},{\cal A}_l\right\}s^i\right)G_2
\]

Now when everything has been computed, one can collect all the terms.
Many terms cancel each other (for example, $I_{3ab}$ cancels with
$I_{1b}$), and the equation for $\tilde{G}$ has the form,
\[
  -ip^\mu{\cal D}_\mu\tilde{G}_{ij}+p^\mu{\cal F}_{\mu\nu}
  {\partial G_s\over\partial p_\nu}
  \left(\delta^{ij}-{p^ip^j\over{\bf p}^2}\right)-
  i{\cal D}_l{\cal F}_{ij}{\partial G_s\over\partial p^l}+
  {ip^k\over2{\bf p}^2}\left({\cal D}_l{\cal F}_{ik}p^j-
  {\cal D}_l{\cal F}_{jk}p^i\right){\partial G_s\over\partial p^l}+
\]
\begin{equation}
  +{ip^k\over2{\bf p}^2}{\cal D}_k{\cal F}_{ij}G_s-
  {i\over{\bf p}^4}p^kp^l\left({\cal D}_l{\cal F}_{ik}p^j-
  {\cal D}_l{\cal F}_{jk}p^i\right)G_s=O(g^8)
  \label{complicated_eq}
\end{equation}
Notice that in the first term in the LHS of Eq.\ 
(\ref{complicated_eq}), one can write $p^k{\cal D}_k\tilde{G}$ instead
of $p^\mu{\cal D}_\mu\tilde{G}$, since $D_0$ is much smaller than
$D_k$.  However, we will leave it as written in Eq.\ 
(\ref{complicated_eq}) for future convenience.  It can be check that
the solution to this equation, up to the order of $g^6$, can be
written in the form,
\begin{equation}
  \tilde{G}^{ij}=\tilde{G}^{\text{kin}}
  \left(\delta^{ij}-{p^ip^j\over{\bf p}^2}\right)
  -{\cal F}_{ij}\left(
  {1\over|{\bf p}|}{\partial G_s\over\partial|{\bf p}|}-
  {G_s\over2{\bf p}^2}\right)+
  {p^k\over{\bf p}^2}\left({\cal F}_{ik}p^j-{\cal F}_{jk}p^i\right)
  \left({1\over2|{\bf p}|}{\partial G_s\over\partial|{\bf p}|}
  -{G_s\over{\bf p}^2}
  \right)
  \label{GGkin}
\end{equation}
where we have introduced $\tilde{G}_{\text{kin}}$ which satisfies the
following equation,
\begin{equation}
  -ip^\mu{\cal D}_\mu\tilde{G}_{\text{kin}}-p^\mu{\cal F}_{\mu\nu}
  {\partial G_s\over\partial p_\nu}=0
  \label{Gkin}
\end{equation}
and made use of the fact that $G_s$ depends on ${\bf p}$ via its
module.  One can recognize in Eq.\ (\ref{Gkin}) the covariant form of
the non-Abelian Vlasov equation.  Making use of the explicit form of
$G_s$, Eq.\ (\ref{Gs}) one can re-express $G^{\text{kin}}$ in term of
the new variable $\delta n^a_{\bf p}(X)$,
\begin{equation}
  \tilde{G}_{ab}^{\text{kin}}(X,p)=-if^{abc}\delta n^c_{\bf p}(X)
  2\pi\delta(\omega^2-{\bf p}^2)
  \label{Gkin_deltan}
\end{equation}
where $\delta n^a_{\bf p}$ satisfies the Vlasov equation in the
non-covariant form,
\begin{equation}
  v^\mu D_\mu\delta n^a_{\bf p}+
  g{\bf vE}^a{\partial n_{\bf p}\over\partial|{\bf p}|}=0
  \label{noncov_Vlasov}
\end{equation}
where $n_{\bf p}$ is the Bose-Einstein distribution function.  Since
$E^a\sim g^5T$, one finds that $\delta n_{\bf p}\sim g^4$.  Eq.\ 
(\ref{noncov_Vlasov}) can be solved explicitly.  The solution reads,
\begin{equation}
  \delta n^a_{\bf p}(t,{\bf x})=
  -g{\partial n_{\bf p}\over\partial|{\bf p}|}
  \int\limits_0^\infty\!du\,U^{ab}(t,{\bf x};t-u,{\bf x}-u{\bf v})
  {\bf vE}^b(t-u,{\bf x}-u{\bf v})
  \label{deltanE}
\end{equation}
Eqs.(\ref{Gkin},\ref{Gkin_deltan},\ref{deltanE}) completely determines
the tree level $G(x,y)$ up to the order of $g^4$.

\subsection{Suppression of loop corrections}

In the previous section we have found tree-level $G$ up to the order
of $g^4$ on any background field $A$ (provided $A\sim g^2T$).  We have
seen that some parts of the propagator can be attributed to the gauge
invariance (the first term in Eq.\ (\ref{G=G0U})) and the deviation of
the distribution function from thermal equilibrium ($\delta n$).
Since small contributions, up to the order of $g^4$, are important for
computing the damping term $j$ in the stochastic equation, a question
arises on the role of loop contributions to $G$: whether there is any
loop diagram that contributes to the thermal propagator (and $j$) to
the order of $g^2$ or $g^4$.  It is known \cite{CalzettaHu} that in
certain cases the inclusion of loops leads to the Boltzmann collision
term in the kinetic equation for $\delta n$.  So, the question we want
to address is essentially whether the collision between hard particles
change the dynamics of modes with $g^2T$ spatial momentum.  We will
present here a set of arguments showing that the corrections coming
from the loops are are much smaller than the order of $g$ we are
interested in and can be neglected.

In fact, according to the general formalism \cite{LifshitzPitaevskii},
when one includes loops, Eq.\ (\ref{collisionless}) will obtains a
collision term in the RHS
\begin{equation}
  I_1+I_2+I_3 = I_{\text{col}}
  \label{collision}
\end{equation}
where
\[
  I_{\text{col}}(x,y) \sim \int\!dz\, (\Sigma(x,z)G(z,y)-
  G(x,z)\Sigma(z,y))
\]
where $\Sigma$ is the self-energy of the hard field.  Since we are
interested up to order $g^6$ in Eq.\ (\ref{collision}), we need to
compute $\Sigma$ up to this order, i.e.\ up to the three-loop level.

To the one-loop level (Fig.\ \ref{fig:1loop}), it is well-known that
there is no contribution to $I_{\text{col}}$ in the absence of
background.  This is because there is a hard particle cannot decay
into two particles due to kinematic constraints.  In the presence of
the background, the analysis is more involved due to the complicated
form of the propagator.  However, essentially the same argument shows
that there is no one-loop contribution to $I_{\text{col}}$.

Two-loop graphs (for example, one in Fig.\ (\ref{fig:2loop}) where one
of the internal propagator may be of $G^{00}$ which corresponds to
exchange of longitudinal gluon) are the lowest ones that generate the
collision term in the absence of background.  To find the same diagram
on the background, one notice that the leading order of magnitude of
the diagram is $g^4$, so one needs to compute these graphs to the
leading and next-to-leading order.  The propagator, up to the
next-to-leading ($g^2$) order can be written as the that in the
absence of background multiplied by the Wilson line,
$G(x,y)=U(x,y)G_0(x,y)$, (we have shown that for spatial components of
$G$, in fact it is also true for $G^{00}$).  Now one can evaluate the
graph in Fig.\ (\ref{fig:2loop}) by replacing each propagator line by
$UG_0$ and the vertices by $t^aD$.  As the result $\Sigma(x,y)$ on the
background differs from the same quantity in the absence of the
background by the Wilson line,
\[
  \Sigma(x,y)=U(x,y)\Sigma(x,y)|_{A=0}+O(g^4)
\]
(the $O(g^4)$ corrections, roughly speaking, comes from the fact that
$U(x,z)U(z,y)$, where $z$ is one of the internal vertices in Fig.\ 
\ref{fig:2loop} is not equal to $U(x,y)$.  However, the difference is
of order $g^4$ which can be neglected).  From that, one can show that
the collision term $I_{\text{col}}$ is the same as of on the absence
of the background except for a Wilson line.  However, it is known that
the collision integral identically vanishes in thermal equilibrium
(notice that $G_0$ is the propagator in thermal equilibrium).  In
particular, the two-loop contributions to the collision integral also
vanishes in the absence of background, so the same is valid when
background is present.

Three-loop graphs are even simpler: they already have the marginal
smallness $g^6$, and to the leading order they can be computed in the
absence of the background.  Invoking the same argument about the
vanishing collision integral in thermal equilibrium, we see that
three-loop graphs have no contribution to the order of $g^6$.

Therefore, the propagator that has been found at tree level has no
correction to the $g^4$ order from loops.

There is a simple (but not rigorous) way to understand the suppression
of loop corrections.  The collision integral $I_{\text{col}}$ is
proportional to $\delta n/\tau$, where $\tau$ is the relaxation time
of the hard modes.  Since $\delta n\sim g^4$, and the relaxation time
for hard modes is of order $1/g^4T$, the collision integral contains
$g^8$, which is much smaller than the order we are interested in
(i.e.\ $g^6$).

\section{Propagators at large spatial separations}
\label{app:propagator2}

To compute the noise kernel $N_{ab}^{ij}(x,y)$, one needs to find the
propagator $G(x,y)$ when $x$ and $y$ are separated by a distance of
order $(g^2T)^{-1}$.  In contrast with the computation of the damping
term $j$, here one needs to compute $G$ only to the leading order.

The propagator can be found most easily by making use of the Fourier
expansion of the operator $a^i$.  Recall that in the absence of the
background field ($A_\mu$=0), one can decompose the field operator
$a^i$ into Fourier components as follows,
\begin{equation}
  a^{ai}(t,{\bf x})=\int\!{d{\bf p}\over(2\pi)^32|{\bf p}|}\left(
  \text{e}^{-i|{\bf p}|t+i{\bf px}}\epsilon_\alpha^i({\bf p})
  b^{a\alpha}_{\bf p}
  +\text{e}^{i|{\bf p}|t-i{\bf px}}\epsilon^{i*}_\alpha({\bf p})
  b^{a\alpha\dagger}_{\bf p}\right)
  \label{aidecomp}
\end{equation}
where $\alpha=1,2$, $\epsilon_\alpha^i({\bf p})$ are the two
transverse polarizations of $a^i$ that are perpendicular to ${\bf p}$,
and $b^{a\alpha}$ and $b^{a\alpha\dagger}$ are annihilation and
creation operators for gauge bosons with polarization $\alpha$ and
color $a$.

In the presence of the background field $A_\mu$, the equation for
$a^i$, up to contribution of order $g^2$ (we need to include, beside
the leading term $O(1)$, also terms of order $O(g^2)$, since the
sub-leading, $O(g^2)$ corrections may accumulate over the distance of
order $O((g^2T)^{-1})$ to a large contribution), can be written as
\[
  D_x^2a^i(x)=0
\]
(cf.\ Eq.\ (\ref{D2ai}), note that $F\sim g^4$).  Eq.\ 
(\ref{aidecomp}) is now modified to
\begin{equation}
  a^{ai}(t,{\bf x})=\int\!{d{\bf p}\over(2\pi)^32|{\bf p}|}\left(
  f^{ab}_{\bf p}(t,{\bf x})\text{e}^{-i|{\bf p}|t+i{\bf px}}
  \epsilon_\alpha^i({\bf p})
  b^{b\alpha}_{\bf p}
  +f^{ab}_{\bf p}(t,{\bf x})\text{e}^{i|{\bf p}|t-i{\bf px}}
  \epsilon^{i*}_\alpha({\bf p})
  b^{b\alpha\dagger}_{\bf p}\right)
  \label{modified}
\end{equation}
where $f^{ab}_{\bf p}(t,{\bf x})$ satisfies the wave equation,
\begin{equation}
  D^2\Bigl(f_{\bf p}(t,{\bf x})
  \text{e}^{-i|{\bf p}|t+i{\bf px}}\Bigr)=0
  \label{wave}
\end{equation}
To specify the boundary condition for $f$, one assumes that when $t<0$
the background field is turned off, $A=0$, and demand that
\[
  f^{ab}_{\bf p}(t,{\bf x})=\delta^{ab}\qquad t<0
\]
Let us check explicitly that, with the required accuracy, the
solution to Eq.\ (\ref{wave}) is
\[
  f^{ab}_{\bf p}(t,{\bf x})=U^{ab}(t,{\bf x};0,{\bf x}-{\bf v}t)
\]
where $U$ is the Wilson line.  Indeed, we are interested in the
leading and $O(g^2)$ contributions to the equations, so
\[
  D_\mu(U\text{e}^{-ipx})
  =-ip_\mu\text{e}^{-ipx}U(t,{\bf x})+
  \text{e}^{-ipx}D_\mu U
\]
where the first term is the leading contribution, while the second
term, containing one derivative of a slowly-varying $U$, is of order
$O(g^2)$.  Taking one more derivative, one finds,
\[
  D^2(U\text{e}^{-ipx})=
  -p^2\text{e}^{-ipx}U-2ip^\mu D_\mu U\text{e}^{-ipx}
\]
where we have neglected the contributions like with two derivatives of
$U$ since it is of order $O(g^4)$.  Now, as we have $p^2=0$ and $p^\mu
D_\mu U=0$, $f$ satisfies the wave equation Eq.\ (\ref{wave}) with the
accuracy that we are interested in, i.e.\ up to $O(g^2)$ corrections
inclusively.

Now making use of the fact that, in thermal equilibrium, $\langle
b^\dagger_{\bf p}b_{{\bf p}'}\rangle=(2\pi)^32|{\bf p}|n_{\bf
  p}\delta({\bf p}-{\bf p}')$, where $n_{\bf p}$ is the Bose-Einstein
distribution function, the propagator can be written in the form,
\begin{eqnarray}
  G^{ij}_{ab}(x,y)= &
  \int\!{d{\bf p}\over(2\pi)^32|{\bf p}|}\,
  \left(f_{\bf p}(x)f_{\bf p}^{-1}(y)\right)^{ab}
  \left(\delta^{ij}-{p^ip^j\over{\bf p}^2}\right)
  \Bigl( & n_{\bf p}\text{e}^{-i|{\bf p}|(t-t')+
  i{\bf p}({\bf x}-{\bf y})}+ \nonumber \\
  & & +(1+n_{\bf p})\text{e}^{i|{\bf p}|(t-t')-
  i{\bf p}({\bf x}-{\bf y})}\Bigr)
  \label{delta_n}
\end{eqnarray}

\section{One-dimensional Langevin equation with noise and damping
  dependent of coordinate: asymmetric discretizations}
\label{app:Langevin_asym}

In this appendix we will consider the following one-dimensional
stochastic equation
\begin{equation}
  \gamma(x)\dot{x}+V'(x)=\xi
  \label{Lang_1}
\end{equation}
\begin{equation}
  \langle\xi(t)\xi(t')\rangle=2T\gamma(x)\delta(t-t')
  \label{Langevin}
\end{equation}
Let us first neglect the $V'$ term in Eq.\ (\ref{Lang_1}).  Let us
first choose the simplest discretization of Eqs.\ 
(\ref{Lang_1},\ref{Langevin}).  Consider the following
finite-difference equation,
\[
  \gamma(x(t)){x(t+\Delta t)-x(t)\over\Delta t}=\xi(t)
\]
\[
  \langle\xi(t)\xi(t+n\Delta t)\rangle={2T\gamma(x(t))\over\Delta t}
\]
Suppose at time moment $t$ the particle's coordinate is $y$, then the
probability of finding, at time moment $t+\Delta t$, the particle at
coordinate $x$ is
\[
  P(t,y\to t+\Delta t,x)=\sqrt{\gamma(y)\over4\pi T\Delta t}\exp
  \left(-{\gamma(y)\over4T\Delta t}(y-x)^2\right)
\]
Thus, the evolution of $\rho(t,x)$ is described by the following
equation,
\begin{equation}
  \rho(t+\Delta t,x)=\int\!dy\,\rho(t,y)
  \sqrt{{\gamma(y)\over4\pi T\Delta t}}\exp\left(
  -{\gamma(y)\over4T\Delta t}(y-x)^2\right)
  \label{naive1discr}
\end{equation}
Now one notes that when $\Delta t$ is small, the integration in Eq.\ 
(\ref{naive1discr}) is saturated mostly by the region of small $y-x$.
Therefore, one can expand the integrand in Eq.\ (\ref{naive1discr})
around $y=x$ and take the integral.  One finds, that in the limit of
$\Delta t\to0$, the evolution of $\rho(t,x)$ follows the following
partial differential equation (the Fokker-Planck equation),
\[
  {\partial\over\partial t}\rho(t,x)=
  T{\partial^2\over\partial x^2}\left(
  {1\over\gamma(x)}\rho(t,x)\right)
\]
When the $V'$ term in Eq.\ (\ref{Lang_1}) is taken into account, the
Fokker-Planck becomes,
\begin{equation}
  {\partial\over\partial t}\rho(t,x)=
  T{\partial^2\over\partial x^2}\left(
  {1\over\gamma(x)}\rho(t,x)\right)+
  {\partial\over\partial x}
  \left({V'(x)\over\gamma(x)}\rho(t,x)\right)
  \label{FP1d1_app}
\end{equation}

Now let us choose another discretization of the Langevin
equation.  Instead of taking the value of $\gamma$ at $x(t)$, let us
take its value at $x(t+\Delta t)$.  In other words, consider
\begin{equation}
  \gamma(x(t+\Delta t)){x(t+\Delta t)-x(t)\over\Delta t}=\xi(t)
  \label{discr2_1}
\end{equation}
\begin{equation}
  \langle\xi(t)\xi(t+n\Delta t)\rangle=
  {2\gamma(x(t+\Delta t))\over\Delta t}\delta_{n0}
  \label{discr2_2}
\end{equation}
Eqs.\ (\ref{discr2_1},\ref{discr2_2}) may present some difficulty for
their understanding: to generate the noise at the time moment $t$,
according to Eq.\ (\ref{discr2_2}), one needs to know the coordinate
at the time moment $t+\Delta t$.  We will understand Eqs.\ 
(\ref{discr2_1},\ref{discr2_2}) as we did for Eqs.\ 
(\ref{Langevin_gen},\ref{noise_gen}), i.e.\ by considering a
statistical ensemble of all possible path $x(t)$ where each path is
associated with a weight,
\[
  \rho[x(t)]\sim\exp\left(-\sum_{n=-\infty}^\infty
  {\Delta t\over4T\gamma(n\Delta t)}\left[\gamma(n\Delta t)
  {x(n\Delta t)-x((n-1)\Delta t)\over\Delta t}\right]^2\right)
\]
In this understanding, the probability of going from the point $y$ at
$t$ to the point $x$ at time $t+\Delta t$ is
\begin{equation}
  P(t,y\to t+\Delta t,x)=C\exp\left(-{\gamma(x)\over4T\Delta t}(y-x)^2
  \right)
  \label{P1}
\end{equation}
where $C$ is the normalization constant defined by the condition of
probability conservation,
\[
  \int\!dx\,P(t,y\to t+\Delta t,x)=1
\]
The Fokker-Planck equation can be derived similarly with the previous
method of discretization, now the result reads,
\begin{equation}
  {\partial\rho\over\partial t}=T{\partial\over\partial x}
  \left({1\over\gamma^3}{\partial\over\partial x}
  \left(\gamma^2\rho\right)\right)+
  {\partial\over\partial x}\left({V'\over\gamma}\rho\right)
  \label{FP1d2_app}
\end{equation}
where the contribution from the term $V'$ in Eq.\ (\ref{Lang_1}) has
been also taken into account.  It is clear that this equation is
different from Eq.\ (\ref{FP1d1_app}), derived using another scheme of
discretization.  Therefore, we see that the motion of the particle is
not uniquely defined by Eqs.\ (\ref{Lang_1},\ref{Langevin}).

\section{One-dimensional Langevin equation with noise and damping
  dependent of coordinate: a symmetric discretization}
\label{app:Langevin}

In this Appendix we will consider the same stochastic equation as in
Appendix \ref{app:Langevin_asym}, Eqs.\ (\ref{Lang_1},\ref{Langevin}).
In Appendix \ref{app:Langevin_asym} we have seen that the
Fokker-Planck equation depends on the the way one ``regularizes'' the
stochastic dynamics.  Here we will use the following prescription.  We
add a small second time derivative term to the LHS of Eq.\ 
(\ref{Lang_1}),
\[
  \epsilon\ddot{x}+\gamma\dot{x}+V'(x)=\xi(t)
\]
and understand this equation and Eq.\ (\ref{Langevin}) as the limit
$\Delta\to0$ of the following finite-difference equations,
\begin{equation}
  \epsilon{x(t+\Delta t)-2x(t)+x(t-\Delta t)\over\Delta^2}+
  \gamma(x(t)){x(t+\Delta t)-x(t-\Delta t)\over2\Delta t}
  +V(x(t))=\xi(t)
  \label{discretized}
\end{equation}
\begin{equation}
  \langle\xi(t)\xi(t+n\Delta t)\rangle=
  {2T\over\Delta t}\gamma(x(t))\delta_{n0}
  \label{discretized_noise}
\end{equation}
Notice that $\dot{x}$ is approximated with the accuracy of $(\Delta
t)^2$, in contrast with the prescriptions in Appendix
\ref{app:Langevin_asym} where the accuracy is only $\Delta t$.  The
limit $\Delta\to0$ should be taken {\em before} the limit
$\epsilon\to0$.  In other words, we assume that $\Delta\ll\epsilon$,
while both are small.  It is useful to introduce the momentum,
\[
  p(t)={x(t)-x(t-\Delta t)\over\Delta t}
\]
Eq.\ (\ref{discretized}) now can be recasted into the form,
\[
  \left({\epsilon\over\Delta t}+
  {\gamma(x(t))\over2}\right)p(t+\Delta t)-
  \left({\epsilon\over\Delta t}-{\gamma(x(t))\over2}\right)p(t)+
  V'(x(t))=\xi
\]
Neglecting small corrections, the evolution in the $(x,p)$ space is
described by the following couple of equations,
\begin{eqnarray}
  x(t+\Delta t) & = & x(t)+p(t+\Delta t)\Delta t \nonumber \\
  p(t+\Delta t) & = &
  \left(1-{\gamma(x(t))\over\epsilon}\Delta t\right)p(t)-
  {\Delta t\over\epsilon}V'(x(t))+{\Delta t\over\epsilon}\xi(t)
  \label{disc_1or}
\end{eqnarray}
These equations, together with Eq.\ (\ref{discretized_noise})
completely defines the evolution of the stochastic system.  The state
of the system at any time slice $t$ is described by a couple of
variable $(x,p)$.  For a given state, one generates a Gaussian noise
$\xi$ having the mean square equal to $2T\gamma(x(t))/\Delta t$ and
run the system one time step forward using Eqs.\ (\ref{disc_1or}).

Our strategy is as follows.  Denoting by $\rho(t,x,p)$ the probability
distribution of the particle in the $(x,p)$ at time moment $t$, we
will find the equation describing the evolution of $\rho(t,x,p)$ with
time.  After that, we will see that at time scales large compared to
$\epsilon$ this equation reduced to the equation for the distribution
function in the $x$-space alone, $\rho(t,x)=\int\!dp\,\rho(t,x,p)$.

Assuming $\xi$ is Gaussian, from Eqs.\ (\ref{disc_1or}) one finds,
\begin{eqnarray*}
  \rho(t+\Delta t,x,p) & = & \int\!dy\,dq\,\rho(t,y,q)\delta(x-y-p\Delta t)
  {\epsilon\over\sqrt{4\pi T\gamma\Delta t}}\times \\
  & & \times\exp\left[-{\epsilon^2\over4T\gamma\Delta t}\left(
  p-\left(1-{\gamma\over\epsilon}\Delta t\right)q+
  {\Delta\over\epsilon}V'(x)\right)^2\right]=
\end{eqnarray*}
\[
  =\int\!dy\,dq\,\rho(t,x-p\Delta t,q)
  {\epsilon\over\sqrt{4\pi T\gamma\Delta t}}
  \exp\left[-{\epsilon^2\over4T\gamma\Delta t}\left(
  p-\left(1-{\gamma\over\epsilon}\Delta t\right)q+
  {\Delta\over\epsilon}V'(x)\right)^2\right]
\]
In the limit $\Delta t\to0$, the integral is saturated by the
integration region where $q$ is closed to $p$, where one can replace
$\rho(q)$ by its Taylor expansion around $p$,
$\rho(q)=\rho(p)+(q-p)\rho'(p)+{1\over2}(q-p)^2\rho''(p)+\cdots$.  In
this limit, one finds that $\rho(t,x,p)$ satisfies the following
diffusion equation,
\begin{equation}
  {\partial\rho\over\partial t}=
  {T\gamma\over\epsilon^2}{\partial^2\rho\over\partial p^2}-
  p{\partial\rho\over\partial x}+
  {1\over\epsilon}(\gamma p+V'){\partial\rho\over\partial p}+
  {\gamma\over\epsilon}\rho
  \label{diff_xp}
\end{equation}
In all our discussions above we have kept $\epsilon$ finite.  Now let
us take the limit $\epsilon\to0$ and try to find the evolution of the
distribution function on time scale of order 1.  Assuming that the $p$
is of order $\epsilon^{-1/2}$ (we will check this assumption a
posteriori), if one keeps only leading on $\epsilon$ terms in the RHS
of Eq.\ (\ref{diff_xp}), the latter reduces to
\[
  {\partial\rho\over\partial t}=
  {T\gamma\over\epsilon^2}{\partial^2\rho\over\partial p^2}+
  {\gamma\over\epsilon}p{\partial\rho\over\partial p}+
  {\gamma\over\epsilon}\rho
\]
The equation has solutions in the form,
\[
  \rho(t,x,p)=\sqrt{\epsilon\over2\pi T}
   \text{e}^{-{\epsilon p^2/2T}}a(x)
\]
where $a$ is an arbitrary function of $x$.  For this solution the
typical value of $p$ is obviously of order $\epsilon^{-1/2}$.  Now,
let us look for the solution to Eq.\ (\ref{diff_xp}) in form of the
following series,
\begin{equation}
  \rho(t,x,p)=\rho_0(t,x,p)+\rho_1(t,x,p)+\rho_2(t,x,p)+\cdots
  \label{rho_ansatz}
\end{equation}
where the expansion parameter is $\epsilon^{1/2}$,
$\rho_1/\rho_0\sim\rho_2/\rho_1\sim\epsilon^{1/2}$, and
\begin{equation}
  \rho_0(t,x,p)=\sqrt{\epsilon\over2\pi T}
  \text{e}^{-\epsilon p^2/2T}a(t,x)
  \label{rho0}
\end{equation}
where $a(t,x)$ is a function that varies on the times scale of order
$\epsilon^0$, $\dot{a}/a\sim1$.  Substituting Eq.\ (\ref{rho_ansatz})
to Eq.\ (\ref{diff_xp}) and putting together terms of the same order
on $\epsilon$, one finds the following equations for $a$, $\rho_1$ and
$\rho_2$,
\begin{equation}
  {\gamma\over\epsilon}\left(
  {T\over\epsilon}{\partial^2\rho_1\over\partial p^2}+
  p{\partial\rho_1\over\partial p}+\rho_1\right)+
  \left({V'\over\epsilon}{\partial\rho_0\over\partial p}-
  p{\partial\rho_0\over\partial x}\right)=0
  \label{rho_1_eq}
\end{equation}
\begin{equation}
  \sqrt{\epsilon\over2\pi T}\text{e}^{-\epsilon p^2/2T}
  {\partial a\over\partial t}=
  {\gamma\over\epsilon}\left(
  {T\over\epsilon}{\partial^2\rho_2\over\partial p^2}+
  p{\partial\rho_2\over\partial p}+\rho_2\right)+
  \left({V'\over\epsilon}{\partial\rho_1\over\partial p}-
  p{\partial\rho_1\over\partial x}\right)
  \label{a_equation}
\end{equation}
Substituting $\rho_0$ in Eq.\ (\ref{rho0}) to Eq.\ (\ref{rho_1_eq}),
one obtains the following equation,
\[
  {\gamma\over\epsilon}\left(
  {T\over\epsilon}{\partial^2\rho_1\over\partial p^2}+
  p{\partial\rho_1\over\partial p}+\rho_1\right)=
  p\text{e}^{-\epsilon p^2/2T}\left(
  {\partial a\over\partial x}+{V'\over T}a\right)
\]
from which one can find $\rho_1$,
\[
  \rho_1=-{\epsilon\over\gamma}p\text{e}^{-\epsilon p^2/2T}
  \left({\partial a\over\partial x}+{V'\over T}a\right)
\]
Substituting this expression for $\rho_1$ to Eq.\ (\ref{a_equation}),
one find $\rho_2$
\[
  \rho_2={\epsilon^2\over2\gamma}p^2\text{e}^{-\epsilon p^2/2T}
  \left({V'\over\gamma T}\left(a'+{V'\over T}\right)+
  {\partial\over\partial x}\left({1\over\gamma}
  \left(a'+{V'\over T}a\right)\right)\right)
\]
and the equation that $a$ must satisfy,
\[
  {\partial a\over\partial t}=T{\partial\over\partial x}{1\over\gamma}
  \left({\partial a\over\partial x}+{V'\over T}a\right)
\]
According to Eqs.\ (\ref{rho_ansatz}) and (\ref{rho0}), $a$ can be
considered as the distribution function of particle with respect to the
coordinate alone.  In other words, with the accuracy $\epsilon^{1/2}$
\[
  a(t,x)=\int\!dp\,\rho(t,x,p)
\]
Re-denoting $a(t,x)$ as $\rho(t,x)$, we find, finally, the
Fokker-Planck equation,
\begin{equation}
  {\partial\over\partial t}\rho(t,x)={\partial\over\partial x}\left(
  {T\over\gamma(x)}{\partial\over\partial x}\rho(t,x)\right)+
  {\partial\over\partial x}\left({V'(x)\over\gamma(x)}\rho(t,x)\right)
  \label{FP_1d_final}
\end{equation}

\setlength{\unitlength}{1cm}

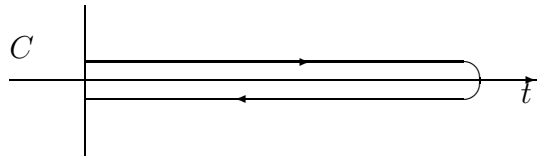
\begin{figure}
\begin{center}
\begin{picture}(7,2)
\put(0,1){\vector(1,0){7}}
\put(1,0){\line(0,1){2}}
\put(1,1.25){\vector(1,0){3}}
\put(4,1.25){\line(1,0){2}}
\put(6,0.75){\vector(-1,0){3}}
\put(3,0.75){\line(-1,0){2}}
\put(6,1){\oval(0.5,0.5)[r]}
\put(6.8,0.7){$t$}
\put(0,1.3){$C$}
\end{picture}
\end{center}
\caption{The Schwinger-Keldysh Close-Time-Path contour}
\label{figure1}
\end{figure}

\vskip 1cm

\begin{figure}
\begin{center}
\leavevmode
\epsfxsize=2in \epsfbox{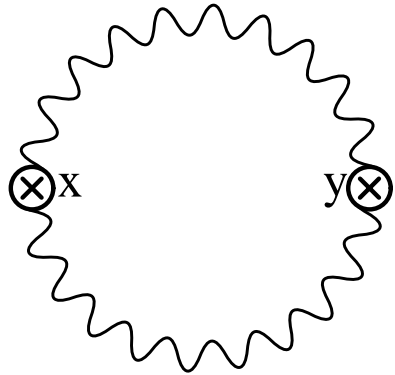}
\end{center}
\caption{A graph with one hard loop}
\label{fig:1loop}
\end{figure}

\vskip 1cm

\begin{figure}
\begin{center}
\leavevmode
\epsfxsize=2in \epsfbox{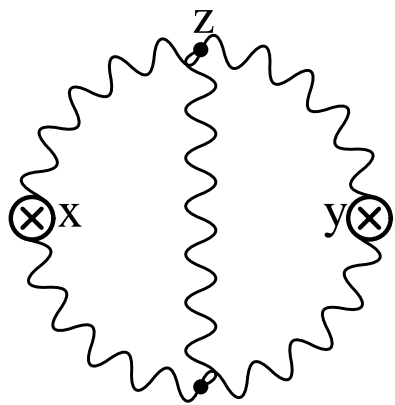}
\end{center}
\caption{A graph with two hard loops}
\label{fig:2loop}
\end{figure}

\end{document}